%% file: main.tex
\documentclass[letterpaper, 10 pt, conference]{ieeeconf}  

\usepackage{graphicx,amsmath}      
\usepackage{amssymb}  
\usepackage{import}
\usepackage{algpseudocode}
\usepackage{tabularx}
\usepackage{soul}
\usepackage{stmaryrd} 
\usepackage{booktabs}
\usepackage{siunitx}
\usepackage{url}
\usepackage{hyperref}

\newtheorem{definition}{Definition}

\IEEEoverridecommandlockouts  
\allowdisplaybreaks 

\newcommand{\BGF}[1]{\textcolor{black}{#1}}
\newcommand{\BG}[1]{\textcolor{black}{#1}}
\newcommand{\BF}[1]{\textcolor{black}{#1}}
\newcommand{\MS}[1]{\textcolor{black}{#1}}
\newcommand{\MAS}[1]{\textcolor{black}{#1}}
\newcommand{\GR}[1]{\textcolor{black}{#1}}
\newcommand{\MADS}[1]{\textcolor{black}{#1}}

\title{\LARGE \bf
Point-to-Cloud NMPC with Smooth Avoidance Constraints$^*$
}

\author{Brener G. Ferreira$^{1}$, Vinicius M. Gonçalves$^{1,2}$, Marcelo A. Santos$^{3}$, Guilherme V. Raffo$^{1,4}$
\thanks{*This manuscript version is made available under the CC-BY-NC-ND 4.0 license. This work was supported by the Brazilian agencies Coordination for the Improvement of Higher Education Personnel (CAPES) – Finance Code 001, CNPq under grants 317058/2023-1 and 422143/2023-5, and FAPEMIG. Marcelo A. Santos acknowledges support from the Lombardy Region under the PR FESR 2021–2027 "Collabora \& Innova" call (Decree no. 11969, 2 August 2024), Project "HARMONY" (CUP: E59I25000850007).}
\thanks{$^{1}$B. G. Ferreira, V. M. Gonçalves and G. V. Raffo are with the Graduate Program in Electrical Engineering, Universidade Federal de Minas Gerais, Belo Horizonte, Brazil. {\tt\small \{brenerufmg, vinicius.mariano, raffo\}@ufmg.br}.}
\thanks{$^{2}$ V. M. Gonçalves is also with the Department of Electrical Engineering, Federal University of Minas Gerais, Belo Horizonte, Brazil.}
\thanks{$^{3}$M. A. Santos is with the Department of Management, Information and Production Engineering, University of Bergamo, Bergamo, Italy. {\tt\small marcelo.santos@unibg.it}.}
\thanks{$^{4}$ G. V. Raffo is also with the Department of Electronic Engineering, Federal University of Minas Gerais, Belo Horizonte, Brazil.}
}      

\newcommand{\commentsymbol}{//}
\algrenewcommand\algorithmiccomment[1]{\ \commentsymbol{} \textit{#1}}

\begin{document}

\maketitle
\thispagestyle{empty}
\pagestyle{empty}

\begin{abstract}
This paper proposes a finite-horizon optimal control strategy for set-point tracking using a nonlinear model predictive control framework with integrated avoidance capabilities. The formulation employs a smooth \BF{point-to-cloud} distance metric that ensures continuously differentiable and numerically \BF{well-conditioned} gradients, even in the presence of regions with complex \GR{and} nonconvex geometries. This smoothness \BF{allows safety constraints to be formulated consistently and differentiably} through \BF{c}ontrol \BF{b}arrier \BF{f}unctions, \BF{resulting in a reliable avoidance behavior for} the closed-loop system. Additionally, stationary artificial \MAS{variables} are introduced \BF{in the optimal control problem to preserve feasibility under changing set-points.} The proposed approach is validated through numerical \GR{experiments} \BF{of an aerial robot}, demonstrating accurate tracking and smooth obstacle avoidance in complex environments.

\end{abstract}


\section{Introduction}

\GR{Nonlinear model predictive control (NMPC)} has emerged as \BF{a} powerful framework for \GR{controlling} dynamic systems subject to constraints, enabling optimal decision-making through a receding horizon strategy\GR{. In this strategy,} the control sequence \BF{is iteratively updated based on the current system information} \cite{camacho2007model}. \MAS{While many NMPC formulations are designed to drive systems toward fixed operating points, they are often employed to track piecewise-constant set-points, where transitions between set-points may lead to feasibility loss \cite{limon2018nonlinear}.} 
To address this limitation, tracking NMPC formulations have been developed by embedding \MAS{artificial variables} within the optimization problem, ensuring \BF{feasibility and asymptotic convergence \MAS{under} changing references}.

A critical challenge in NMPC arises when \BF{the avoidance of complex and nonconvex \GR{unfeasible} regions} must be enforced during tracking. Traditional formulations rely on point-wise Euclidean distance functions as safety measures. However, \MAS{this} distance is non-differentiable in regions of geometric ambiguity, such as sharp corners, nonconvex boundaries, and configurations equidistant to multiple \BF{region} points \cite{Albano2022}. These non-differentiabilities compromise gradient-based numerical solvers and hinder the \MAS{formulation of smooth constraints defined in terms of distance.} 

\BF{Among the various approaches to integrate avoidance constraints into NMPC, such as hard constraints \cite{mayne2000constrained}, soft penalties \cite{hermans2018penalty}, \GR{and} potential-based terms \cite{santos2023nonlinear}, \MAS{control barrier functions (CBFs)} have emerged as a systematic method to guarantee safety by ensuring forward invariance of a desired safe set \cite{ames2016control, Zhang2023}.} They provide a formal mechanism to maintain safe distances from 
\GR{unfeasible} regions, thereby enhancing the safety margin of the closed-loop system. However, existing CBF-based NMPC \MAS{often relies} on classical distance metrics, which limit their smoothness properties and compromise the reliability of the optimization process. 

\BG{Abrupt changes in gradient directions \MAS{can} lead to oscillatory avoidance behavior and poor numerical conditioning. To \MAS{mitigate} these issues, recent works have proposed \MAS{refined} distance metrics that \MAS{enhance} smoothness and responsiveness in \MAS{avoidance-oriented NMPC} frameworks. For \MAS{example}, \cite{11217314} have introduced extended Euclidean metrics that incorporate obstacle dynamics into the safety evaluation process, while \cite{liang2025point} have proposed a point-to-cloud CBF-MPC formulation capable of handling \MAS{obstacles with} complex and irregular geometries. \MAS{Particularly relevant to this work, \cite{goncalves2024smooth} have proposed} smoothed point-to-cloud distance metrics as differentiable surrogates that approximate the Euclidean distance while ensuring well-behaved higher-order derivatives. These metrics preserve essential geometric information and allow explicit tuning between fidelity and numerical regularity, \MAS{a property especially} advantageous for NMPC formulations \MAS{requiring} smooth and stable \MAS{gradients}. Despite their potential, the use of such smoothed distances has not yet been systematically exploited within \MAS{CBF-based} NMPC formulations for tracking \MAS{with} avoidance capabilities.}


\MAS{Therefore, } 
\GR{this work proposes a CBF-based NMPC framework that incorporates a smoothed point-to-cloud distance metric. This formulation significantly improves numerical conditioning and gradient smoothness in the presence of complex and nonconvex unfeasible regions. It also ensures recursive feasibility during set-point changes through the inclusion of stationary artificial variables. \MAS{Thus, the} main contributions \MAS{of this work are twofold}: (i) the introduction of a continuously differentiable point-to-cloud distance metric that enables smooth and numerically stable gradient evaluations; and (ii) the development of a CBF-based tracking NMPC formulation that \MAS{guarantees} both safety and feasibility during the tracking of piecewise-constant references.} 
%
%
The effectiveness of the proposed controller is \BF{corroborated through numerical experiments} \GR{with an aerial robot}.

\MAS{The paper is organized as follows:} Section II describes the \GR{standard} tracking NMPC formulation; Section III presents the \MAS{smoothed distance metric}; \BF{Section IV introduces the CBF-based \MAS{tracking NMPC}}; Section V \GR{provides} \BF{numerical results}; and Section VI \GR{offers the concluding remarks}.

\section{\GR{Standard} tracking NMPC formulation}
\label{sec:tracking_nmpc}

Consider the 
nonlinear dynamical system
\begin{align}
x_{k+1} &= f(x_{k},u_{k}), \label{eq:modelodinamico}
\end{align}
with $x_{k}\in\mathbb{R}^n$ and $u_{k}\in\mathbb{R}^m$ being the state and input vectors, respectively. The state and input pairs are constrained by
\begin{align}
(x_{k},u_{k}) &\in \MS{\mathcal{X}} \times \MS{\mathcal{U}}, \quad \forall k \in \mathbb{I}_{\geq 0},
\label{eq:constraintsStatesInputs}
\end{align}
where $\MS{\mathcal{X}}\subseteq\mathbb{R}^n$ and $\MS{\mathcal{U}}\subseteq\mathbb{R}^m$ are the admissible sets.

Let the system output be given by
\begin{align}
\BF{y_k = h(x_k),} \qquad 
\label{eq:output}
\end{align}
where \BF{$\BGF{y_k} \in \mathbb{R}^p$ typically represents physical quantities relevant for the control objectives.}

To enable \BF{set-point} tracking at the output level, the controller drives the system toward an artificial \BF{equilibrium} $(x_a,u_a)$, \BF{that unequivocally defines the steady output $y_a$}\MS{, as described in \cite{limon2018nonlinear}. Therefore, the} artificial pair \BGF{\MS{must} satisfy}
\begin{align}
x_a = f(x_a,u_a), \qquad (x_a,u_a)\in  \BF{\MS{\mathcal{Z}_s}},
\label{eq:steady_state_artificial}
\end{align}
\BF{where $\MS{\mathcal{Z}_s}$ is the set of admissible equilibria, defined as}
\begin{equation}
\MS{\mathcal{Z}_s} \!=\! \{(x_s,u_s)\!:\! x_s \!=\! f(x_s,u_s),\! (x_s,u_s)\!\in\! \lambda(\MS{\mathcal{X}} \!\times \!\MS{\mathcal{U}})\MS{\}},
\end{equation}
\BF{with $\lambda \in (0,1)$ being a relaxation factor to avoid controllability loss associated with artificial equilibria lying on the boundary of the admissible \MS{space $\mathcal{X} \times \mathcal{U}$ \cite{limon2018nonlinear}}}.

The finite-horizon cost functional is defined as
\begin{align}
\label{eq:trackingCost_final}
V_{N}^{t}(x,y_t;\boldsymbol{u},x_a,u_a)
&= \sum_{j=0}^{N-1} \|x_{j}-x_{a}\|_{Q}^{2} + \|u_{j}-u_{a}\|_{R}^{2} \nonumber\\
&\quad + \|y_{a} - y_{t}\|_{T}^{2},
\end{align}
where \BF{$Q \succeq 0$, \MS{$R \succ 0$, and $T \succ 0$}} are weighting matrices, $N$ \BF{denotes} the prediction horizon, and $\boldsymbol{u}=\{u_0,\dots,u_{N-1}\}$ \BF{is the control sequence}. \BF{In \eqref{eq:trackingCost_final}, the stage cost ensure convergence} of the predicted \BF{evolution of the system \MS{to} the stationary artificial variables} $(x_a,u_a)$, whereas the \MS{stationary offset} term \BF{ensures convergence to the desired set-point} $y_t$.

\BF{Considering model \eqref{eq:modelodinamico} for prediction, constraints \eqref{eq:constraintsStatesInputs} and \eqref{eq:steady_state_artificial}, and the cost functional \eqref{eq:trackingCost_final}, the \GR{standard} tracking NMPC is formulated as \MS{follows:}} 
\begin{subequations}
\label{eq:trackingNMPC_final}
\begin{align}
\min_{\boldsymbol{u},x_a,u_a}\ & V_N^{t}(x,y_t;\boldsymbol{u},x_a,u_a),\nonumber\\
\mathrm{s.t.}\ & x_0 = x, \label{eq:ic}\\
& x_{j+1}=f(x_j,u_j), \quad j\in\mathbb{I}_{0:N-1}, \label{eq:dynamics}\\
& (x_j,u_j)\in \MS{\mathcal{X}}\times \MS{\mathcal{U}}, \quad j\in\mathbb{I}_{0:N-1}, \label{eq:state_input_constraints}\\
& (x_a,u_a)\in \MS{\mathcal{Z}_s},\\
&\BGF{y_a = h(x_a,u_a)}, \\
&x_N=x_a,\label{eq:artificial_constraint}
\end{align}
\end{subequations}
\BF{with constraint \eqref{eq:artificial_constraint} being a relaxed terminal equality constraint added for stability purposes \MS{\cite{limon2018nonlinear}.}}

\section{Smoothed Point-to-\GR{Cloud} Distance}
\label{sec:smoothed_distance}


\BF{This section \MADS{considers the} smooth point-to-cloud distance metric \MADS{introduced in \cite{goncalves2024smooth} and its integration within the NMPC framework} to represent \GR{nonfeasible} regions. \MS{This} metric provides a continuously differentiable ($\mathcal{C}_{\infty}$ class) and numerically stable approximation of the Euclidean distance, ensuring well-behaved gradients and compatibility with CBF-based constraints.}

\subsection{Discrete Smoothed Distance Formulation}

Let $\BF{\mathcal{A}} = \BF{\{a_j\}_{j=1}^{m_{\mathcal{A}}}} \subset \mathbb{R}^p$ be a finite set of sampled points representing a \GR{nonfeasible} region in the output space, and \GR{let} $y \in \mathbb{R}^p$ \GR{be} \MS{the system output}. The smoothed point-to-cloud distance between $y$ and the set $\mathcal{A}$ is defined as
\begin{align}
    D_{\eta,\sigma}^{\mathcal{A}}\!(y)
    \!=\! 
    -\eta^2
    \!\ln\!\!\left(
    \!\!\frac{1}{V_{\sigma}^{\mathcal{A}}}
    \!\!\sum_{j=1}^{m_{\mathcal{A}}}
    \!\left(W_{\sigma}^{\mathcal{A}}(a_j)
    \exp\!\!\left(
    \!{-}\frac{\|y {-} a_j\|^2}{2\eta^2}
    \!\right)
    \!\right)
    \!\!\right)\!,
    \label{eq:smoothed_distance}
\end{align}
\BF{where \MS{$\ln(\cdot)$ and $\exp(\cdot)$ are, respectively, the natural logarithm and the exponential operators,} and $\eta > 0$ is \MS{a} smoothing parameter controlling the trade-off between geometric accuracy and differentiability.} Smaller values of $\eta$ yield a closer approximation to the Euclidean distance but with sharper gradients, whereas larger values provide smoother behavior at the cost of reduced geometric fidelity.

The parameter $\sigma > 0$ defines the regularization scale, which governs the spatial dispersion of the sampled points through the weighting function
\begin{align}
    W_{\sigma}^{\mathcal{A}}(a_j)
    =
    \exp\!\left(
    -\frac{\|a_j - \mathrm{Cen}(\mathcal{A})\|^2}{2\sigma^2}
    \right),
    \label{eq:weight_function}
\end{align}
\BF{\MS{with} the centroid of the set \MS{being} given by}
\begin{align}
    \mathrm{Cen}(\mathcal{A})
    =
    \frac{1}{m_{\mathcal{A}}}
    \sum_{j=1}^{m_{\mathcal{A}}} a_j.
    \label{eq:centroid}
\end{align}
\BF{Finally}, the total weighted volume of the set is defined as
\begin{align}
    V_{\sigma}^{\mathcal{A}}
    =
    \sum_{j=1}^{m_{\mathcal{A}}} W_{\sigma}^{\mathcal{A}}(a_j),
    \label{eq:weighted_volume}
\end{align}
which serves as a normalization factor in \eqref{eq:smoothed_distance}, ensuring dimensional consistency and invariance under affine transformations \BF{of} $y$ and $\mathcal{A}$. \MS{This normalization also improves the numerical conditioning when the metric is used within the optimization framework}. 

The distance function $D_{\eta,\sigma}^{\mathcal{A}}(y)$ is strictly positive and infinitely differentiable for all $y \in \mathbb{R}^p$. Moreover, as $\eta \to 0$, it converges to the classical Euclidean distance under mild regularity conditions on $\mathcal{A}$ \cite{goncalves2024smooth}. 
\BF{The differentiability of this distance makes it suitable for defining smooth barrier \GR{and} penalty terms in NMPC formulations with avoidance constraints.}

\subsection{Smoothed Projection and Gradient}

\BF{The smoothed point-to-cloud distance defined in \eqref{eq:smoothed_distance} allows for the computation of a continuously differentiable projection and gradient.} 
For the set $\mathcal{A}$ and a given point $y$, the smoothed projection $\Pi_{\eta,\sigma}^{\mathcal{A}}(y)$ is defined as the weighted mean of the sampled points\GR{, that is,}
\begin{align}
    \Pi_{\eta,\sigma}^{\mathcal{A}}(y)
    =
    \frac{
    \sum_{j=1}^{m_{\mathcal{A}}}
    W_{\sigma}^{\mathcal{A}}(a_j)
    a_j
    \exp\!\!\left(
    -\frac{\|y {-} a_j\|^2}{2\eta^2}
    \right)
    }{
    \sum_{j=1}^{m_{\mathcal{A}}}
    W_{\sigma}^{\mathcal{A}}(a_j)
    \exp\!\!\left(
    -\frac{\|y {-} a_j\|^2}{2\eta^2}
    \right)
    }.
    \label{eq:projection}
\end{align}

\BF{This smooth projection can be interpreted as the differentiable counterpart of the classical Euclidean projection onto a discrete set, avoiding non-differentiability issues at boundary transitions.}  
It enables well-defined derivatives everywhere, which is particularly important for gradient-based optimization methods.

The gradient of the smoothed distance with respect to $y$ is given by
\begin{align}
    \nabla_y D_{\eta,\sigma}^{\mathcal{A}}(y)
    =
    y - \Pi_{\eta,\sigma}^{\mathcal{A}}(y),
    \label{eq:gradient}
\end{align}
\BF{where $\nabla_y (\cdot)$ denotes the gradient operator with respect to the output vector $y$.}  
Equation \eqref{eq:gradient} reveals that the gradient direction points from the smoothed projection $\Pi_{\eta,\sigma}^{\mathcal{A}}(y)$ toward the current position $y$. 
\BF{This differentiable property ensures that both the magnitude and direction of the gradient vary continuously, preventing abrupt control actions \MS{when applied in optimal control problems to obtain control actions.}} 

As $\eta \to 0$, the smoothed projection converges to the classical Euclidean projection, and the gradient in \eqref{eq:gradient} approaches the exact direction of minimum distance between $y$ and $\mathcal{A}$.  
\BF{This limit behavior ensures theoretical consistency with the classical Euclidean distance, while maintaining the \MS{aforementioned} numerical advantages}. 
\BF{\MS{Given these properties,} the smoothed distance and its gradient provide a reliable and computationally efficient representation of repulsive interactions between the controlled system and \GR{nonfeasible} regions, \MS{aligning directly with the objectives of the avoidance NMPC formulation developed in this work.}} 

\section{Tracking NMPC with Smoothed Avoidance Constraints}
\label{sec:tracking_cbf}

\BF{This section presents the derivation of smooth \MS{CBFs to define} avoidance constraints within the tracking NMPC framework \cite{SAN25_LCSS,dos2024set}\MS{, where the smoothed distance \eqref{eq:smoothed_distance} is used as a surrogate for the Euclidean distance}. 
Following the penalty-based framework of \cite{ames2016control}, the CBF terms are incorporated into the cost \GR{functional} as smooth penalties rather than imposed as hard constraints, preserving differentiability and maintaining a well-conditioned optimization problem.}

The goal is to ensure that the system outputs remain within a safe region $\mathcal{S} \subset \mathbb{R}^p$ \BF{throughout the prediction horizon, which is defined as}
\begin{align}
\mathcal{S} = \{y \in \mathbb{R}^p \mid \MS{\beta}(y)\geq 0\},
\label{eq:BarrierFunctionL}
\end{align}
where \BF{$\MS{\beta} : \mathbb{R}^p \to \mathbb{R}$} is a continuous function and the interior of $\mathcal{S}$ is assumed to be non-empty.

\begin{definition}[\cite{varun2021motion}, Definition~1]
Given a set $\mathcal{S}$ defined \MS{as} \eqref{eq:BarrierFunctionL}, a function $\MS{\beta}$ is called a CBF if there exist control inputs $u_k \in \mathcal U$ and a constant $\delta \in (0, 1)$ such that, for all $\MAS{y_0} \in \mathcal{S}$,
\begin{equation}
\sup_{u_k \in \MAS{\mathcal{U}}} \MS{\beta}(y_{k+1}) \geq \delta \MS{\beta}(y_k), \MAS{\quad \forall k \ge 0,}
\label{eq:SUP_output}
\end{equation}
\end{definition}
\BF{where $y_{k+1}$ is obtained from the system dynamics} \MS{\eqref{eq:modelodinamico} and \eqref{eq:output} as \GR{$y_{k+1} = h(f (x_k,u_k))$}}.

\BF{Following from \MS{D}efinition 1}, \MAS{the CBF constraint is incorporated within the prediction horizon to ensure safety as}
\begin{align}
\MS{\beta}(y_{j+1}) \geq \delta \MS{\beta}(y_j)
\label{eq:BRF}
\end{align}
can be satisfied for some $\delta \in (0, 1)$, ensuring that $\MS{\beta}(y_j) \geq 0$ for all \MAS{$j$}, as long as $\MS{\beta}(y_0) \geq 0$.

Each \GR{nonfeasible} region is represented in the output space by a finite set of samples \BF{$\mathcal{O}_i = \{o_j\}_{j=1}^{m_O} \subset \mathbb{R}^p$}. 
Considering the smooth distance \eqref{eq:smoothed_distance}, the CBF 
can be designed as
\begin{align}
\MS{\beta}(y,\mathcal{O}_i) =  D_{\eta,\sigma}^{\mathcal{O}_i}(y) - d_{\min},
\label{eq:barrier_dist_output}
\end{align}
where $d_{\min}>0$ is a safety margin.

The violation measure associated with the discrete barrier condition \MS{\eqref{eq:BRF}} is defined as
\begin{align}
g(y_j,\mathcal{O}_i) = (1-\delta)\MS{\beta}(y_j,\mathcal{O}_i) - \MS{\beta}(y_{j+1},\mathcal{O}_i),
\label{eq:g_violation_output}
\end{align}
where $g(y_j,\mathcal{O}_i) \le 0$ implies satisfaction of the safety condition\MS{, while violations $g(y_j,\mathcal{O}_i)>0$ are penalized through the differentiable function}
\begin{align}
F\!\left(y_j,\mathcal{O}_i\right)
= \frac{1}{\kappa}\!\left[\ln\!\left(1 + \exp\!\left(\kappa\, g(y_j,\mathcal{O}_i)\right)\right)\right]^{\epsilon},
\label{eq:F_penalty_output}
\end{align}
\BF{where $\kappa>0$ controls the smoothness (sharpness) and $\epsilon>0$ adjusts the curvature of the \BF{penalty function}. Moreover, as $\kappa \to \infty$, the function $F(y_j,\mathcal{O}_i)$ converges to the penalty $\max\{0, g_i(y_j,\mathcal{O}_i)\}$, recovering the non-smooth behavior of standard formulations.}

\BF{Therefore, by applying the proposed CBF-based penalty to ensure that both the artificial steady output and the predicted system outputs remain outside \GR{nonfeasible} regions, the cost function \eqref{eq:trackingCost_final} is redefined as follows:}
\begin{align}  
&V_{N}(x,y_t,O_i;\boldsymbol{u},x_{a},u_{a}) = \nonumber\\ 
&\sum_{j=0}^{N-1}\!\left(\|x_{j}-x_{a}\|_{Q}^{2}
+\|u_{j}-u_{a}\|_{R}^{2}\right)
+\|y_{a}-y_{t}\|_{T}^{2} \nonumber\\ 
& + \sum_{i=1}^{N_{O}} \left( \mu F(y_a, \mathcal{O}_i) + \!\sum_{j=0}^{N}\!\mu F\!\big(y_j,\mathcal{O}_i)\right),
\label{eq:Custopenalidade_output}
\end{align}
\BF{where $\mu>>0$ is the penalty weight.} 

Finally, incorporating these smooth penalties into the tracking NMPC yields
\begin{align}
\min_{\boldsymbol{u}, x_a, u_a} 
\quad & V_N(x, y_t, \mathcal{O}_i; \boldsymbol{u}, x_a, u_a), \nonumber\\
\mathrm{s.t.} \quad & \eqref{eq:ic}\text{--}\eqref{eq:artificial_constraint}.
\label{eq:formulacaoPenalidade_output}
\end{align}

\BF{This formulation defines the tracking NMPC with smoothed avoidance constraints, enabling safe set-point tracking while ensuring differentiability and improved numerical conditioning.}

\section{Numerical Results}

The nonlinear system adopted in this work corresponds to \GR{a} quadrotor \GR{unmanned aerial vehicle} (UAVs), whose dynamic model was employed to implement the proposed control strategy. Numerical \GR{experiments} were performed in MATLAB R2025a, employing the CasADi toolbox version 3.6.7 \GR{along} with the IPOPT solver to address the nonlinear optimization problem. 

Consider the quadrotor \BF{UAV dynamical model} described in \cite{pereira2021nonlinear}, actuated via \GR{thrust} forces $u = [f_1~ f_2~ f_3~ f_4]^T$, and with states $ x = [x\MS{_\mathcal{I}}~ y\MS{_\mathcal{I}}~ z\MS{_\mathcal{I}}~ \phi~ \theta~ \psi~ \dot{x}~ \dot{y}~ \dot{z}~ \dot{\phi}~ \dot{\theta}~ \dot{\psi}]^T$\MS{, where $x_\mathcal{I}$, $y_\mathcal{I}$, $z_\mathcal{I}$ denote the position coordinates expressed in the inertial frame, 
$\phi$, $\theta$, $\psi$ are the Euler angles describing the UAV orientation, and the remaining variables 
represent their respective time derivatives.}
The \MS{complete} nonlinear dynamics, considering the parameters of Table \ref{tab:quadrotor_params}, are numerically integrated in MATLAB using ODE45.  
\begin{table}[h!]
\centering
\caption{Quadrotor UAV Parameters.}
\label{tab:quadrotor_params}
\footnotesize
\begin{tabularx}{\linewidth}{l c c}
\toprule
\textbf{Description} & \textbf{Symbol} & \textbf{Value} \\
\midrule
Mass & $m$ & 2.24 [kg] \\
Distance to rotors & $h$ & 0.332 [m] \\
Gravity & $g$ & 9.81 [m$\cdot$s$^{2}$] \\
Thrust coefficient & $b$ & $9.5 \times 10^{-6}$ [N$\cdot$s$^2$] \\
Drag coefficient & $k_\tau$ & $1.7 \times 10^{-7}$ [N$\cdot$m$\cdot$s$^2$] \\
Inertia (X-axis) & $I_{xx}$ & 0.0363 [kg$\cdot$m$^2$] \\
Inertia (Y-axis) & $I_{yy}$ & 0.0363 [kg$\cdot$m$^2$] \\
Inertia (Z-axis) & $I_{zz}$ & 0.0615 [kg$\cdot$m$^2$] \\
\bottomrule
\end{tabularx}
\end{table}

\MS{Consider a \(50\times50\times25\) meters workspace where obstacles are represented as point clouds. 
Since the scenario is simulated, the point clouds are generated using quasi-random low-discrepancy Halton sequences~\cite{halton1960}. 
A virtual 3D LiDAR sensor with a spherical detection radius of \(r_s = 3~\mathrm{m}\), centered at the UAV position, provides obstacle measurements. 
Consequently, at each sampling instant, the UAV only has access to the obstacle points within its sensing range, and the NMPC formulation considers these points to compute the optimal control action.
}

Regarding the NMPC parameters, the horizon length is \(N = 35\), and the sampling time is \(T_s = 0.01 \text{ s}\)\BGF{, selected according to the fastest system dynamics, namely the orientation dynamics.} The weighting matrices for the states and inputs are \(Q = \text{diag}(\{1~ 1~ 1~ 0.1~ 0.1~ 1~ 1~ 1~ 1~ 10~ 10~ 1\})\) and \(R = \text{diag}(\{3~ 3~ 3~ 3\})\), respectively. Furthermore, the weighting matrix for the offset cost functional is $T = \text{diag}(\{1000~ 1000~ 1000\})$\GR{, while} \MS{the avoidance penalty weight is $\mu = 5\times10^4$}. \BGF{The matrices \(Q\), \(R\), and \(T\) are selected such that \((Q, R) \ll T \ll \mu\), prioritizing safety, followed by convergence to the target, and then tracking performance and control effort.} The smoothing and barrier function parameters were set to $\eta=0.3$, $ \sigma=0.8$, \MS{and} $\delta=0.95$. 
In terms of constraints, the sets of admissible inputs and states are given, respectively, by \(U = \{0 \leq f_1, f_2, f_3, f_4 \leq 12\}\) and $X = \{-[25 ~ 25~ 0 ]^T \leq [x~ y~ z] \leq [25~ 25~ 25]^T, -[\pi/6~ \pi/6~ \pi ]^T \leq [\phi~ \theta~ \psi] \leq [\pi/6~ \pi/6~ \pi]^T, -[5~ 5~ 3 ]^T \leq [\dot{x}~ \dot{y}~ \dot{z}] \leq [5~ 5~ 3]^T, -[\pi/2~ \pi/2~ \pi/2 ]^T \leq [\dot{\phi}~ \dot{\theta}~ \dot{\psi}] \leq [\pi/2~ \pi/2~ \pi/2] \}^T$.

In this simulation, the quadrotor UAV is required to autonomously move from its initial configuration \([22~ {-}22~ 0 ]^\top\) to the desired configurations \([{-}12~  18~  11]^\top\!\), and \([{-}21~  18~  0]^\top\!\), \MS{sequentially. The} stopping criteria considered for the simulations are the Euclidean distance between the target and the UAV being less than 0.3 meters.

\begin{figure}[!h]	
    \footnotesize
    \centering{
        \def\svgwidth{1\columnwidth}
        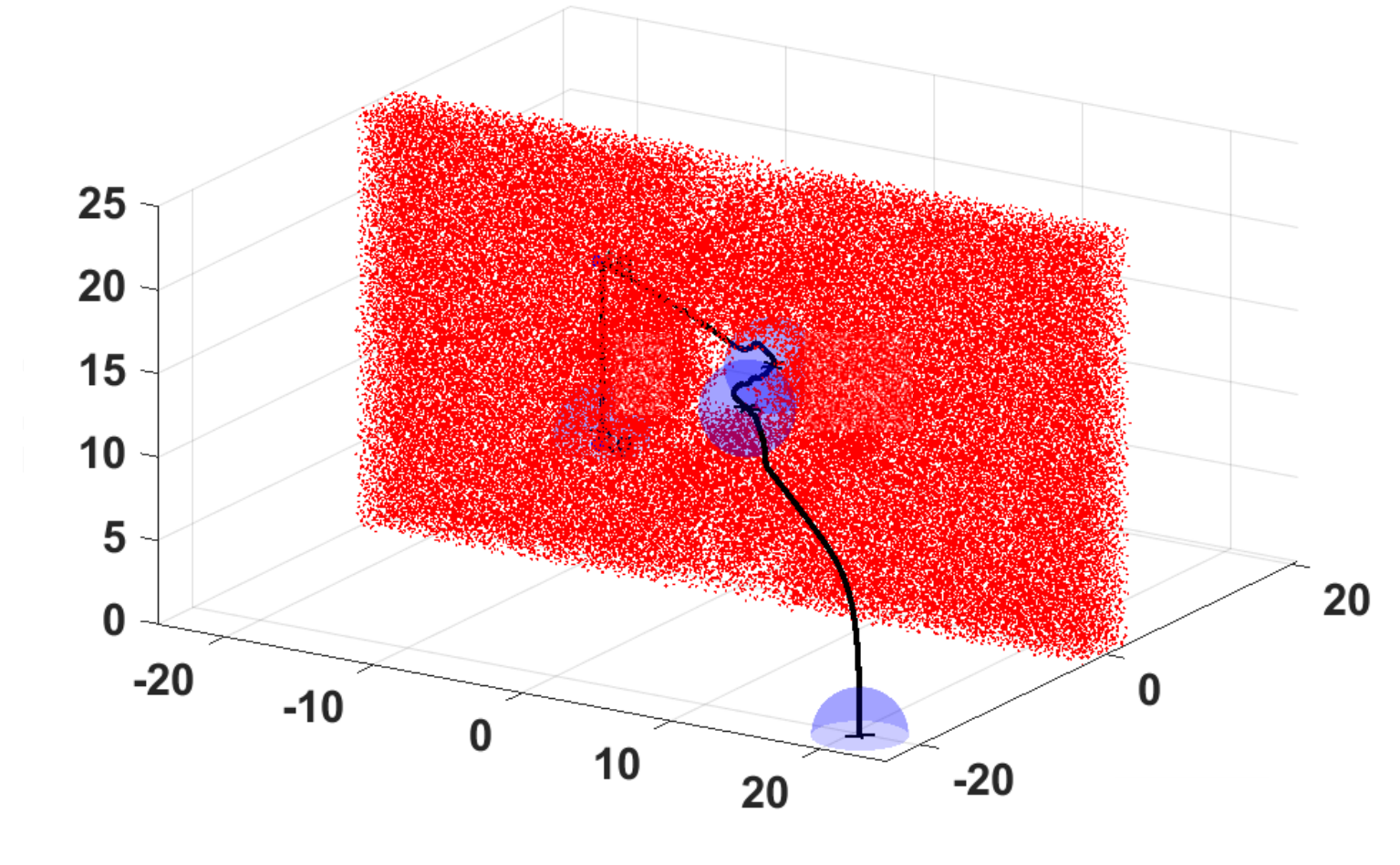}
        \caption{UAV flight trajectory (black line) to safely execute a task while avoiding the nonfeasible region represented as red point clouds. The translucent blue sphere indicates the detection range of the 3D LiDAR.} 
        \label{fig:Obstaculos12}
    \normalsize
\end{figure}

\begin{figure}[!h]	
    \footnotesize
    \centering{
        \def\svgwidth{0.95\columnwidth}
        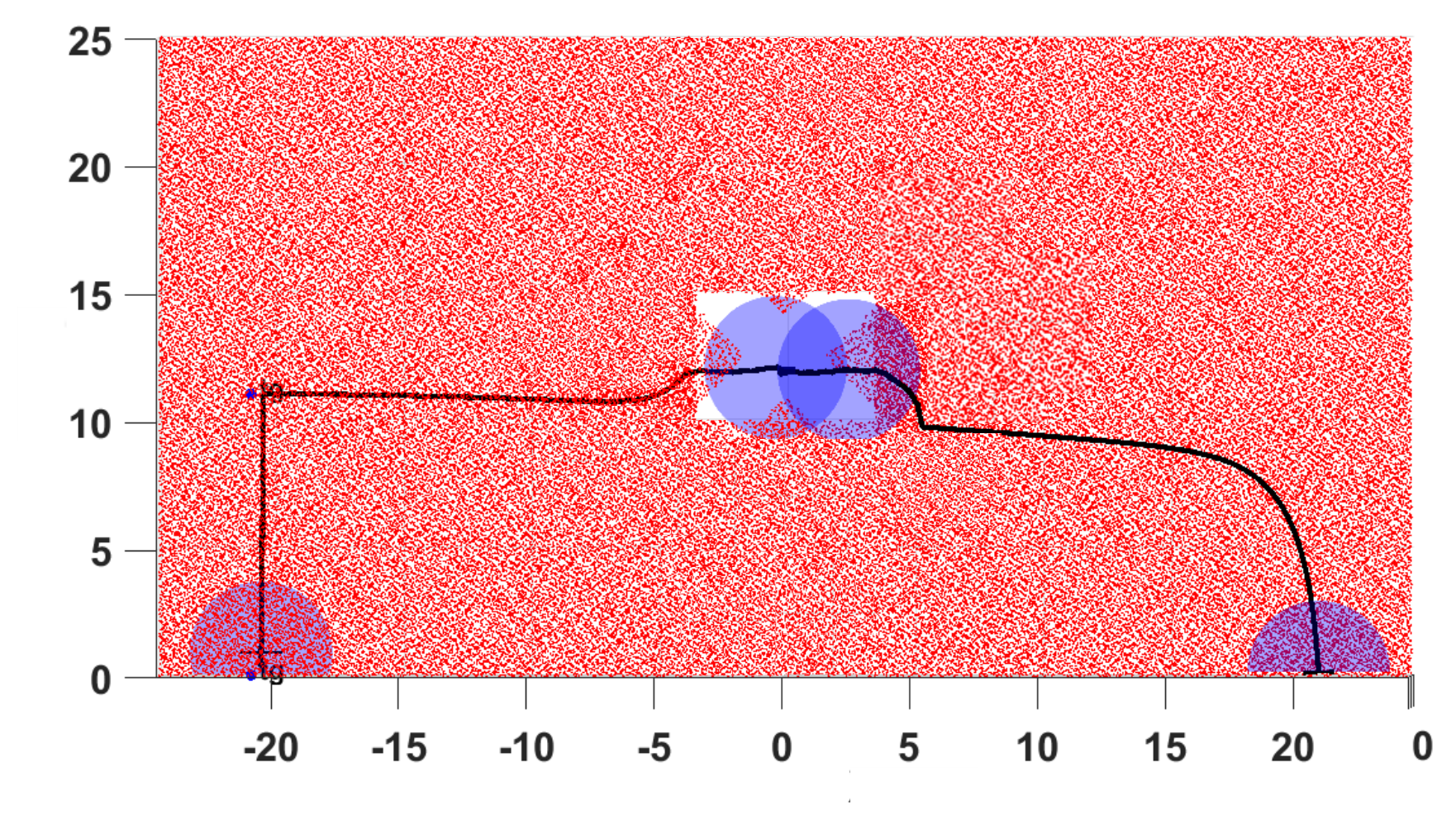}
        \caption{Projection of the trajectory onto
 the XZ-plane.} 
        \label{fig:Obstaculos13}
    \normalsize
\end{figure}

\begin{figure}[!h]	
    \footnotesize
    \centering{
        \def\svgwidth{0.8\columnwidth}
        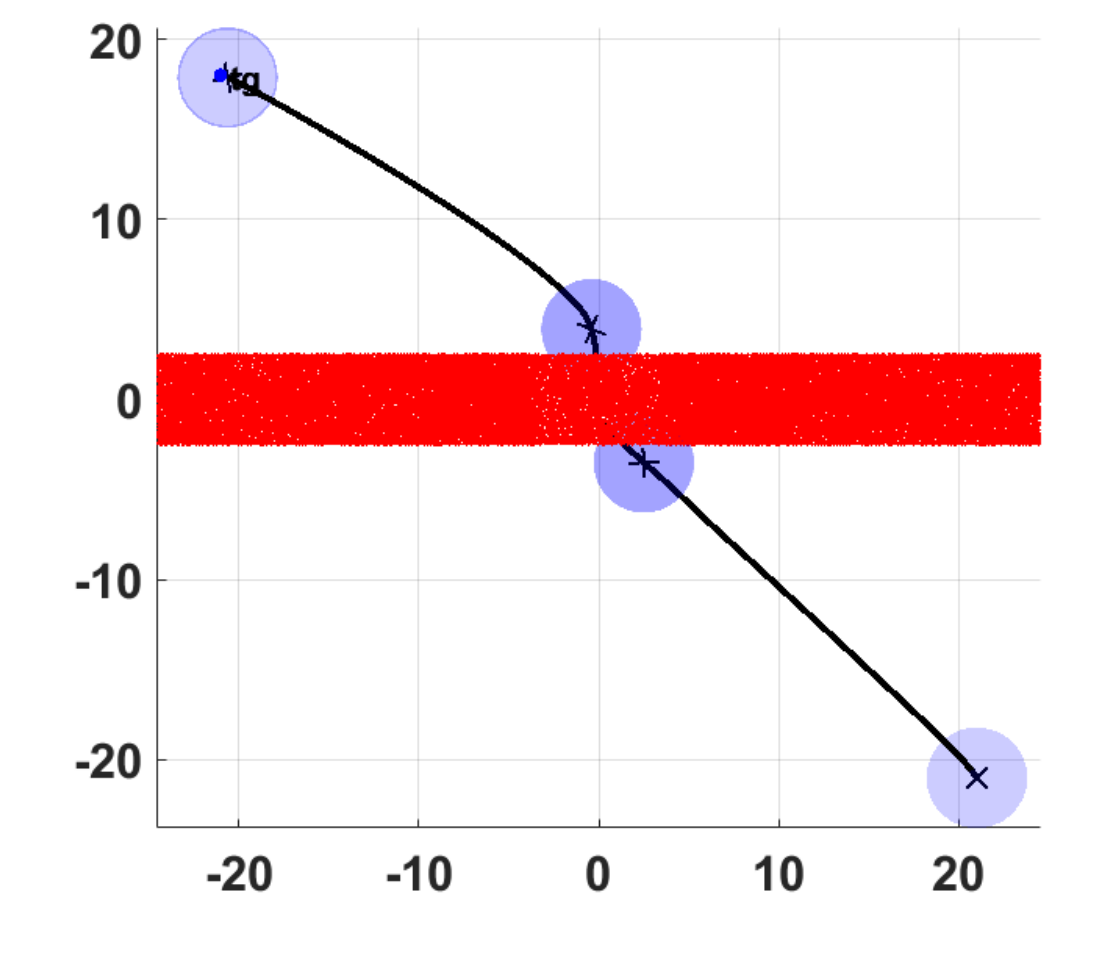}
        \caption{Projection of the trajectory onto
 the XY-plane.} 
        \label{fig:Obstaculos14}
    \normalsize
\end{figure}

\GR{Figs.} \ref{fig:Obstaculos12} \MS{to} \ref{fig:Obstaculos14} \GR{show} that the NMPC drives the UAV toward the reference points while continuously adapting its trajectory to \BF{avoid} \GR{the nonfeasible region}. The black trajectory reshapes smoothly around the red areas, ensuring that penetration never occurs and that a consistent safety margin is preserved, even in close-proximity situations. This smooth adaptation is particularly evident \GR{during} transitions between regions of free space and zones densely occupied by \GR{nonfeasible} regions.

\begin{figure}[!h]	
    \footnotesize
    \centering{
        \def\svgwidth{0.9\columnwidth}
        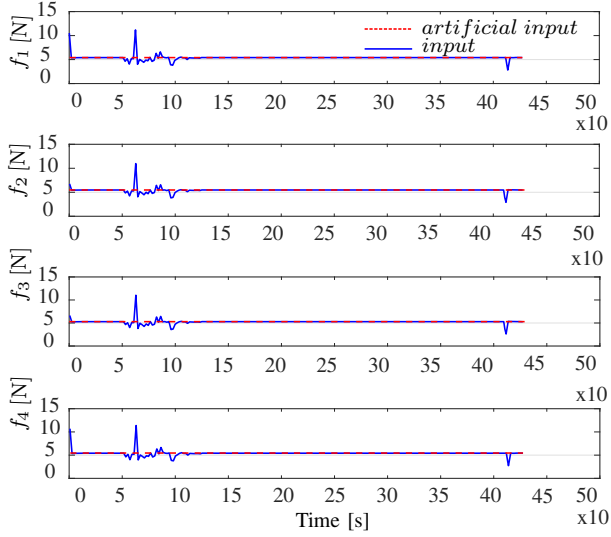}
        \caption{ Time evolution of the thrust lift forces applied to the quadrotor UAV.} 
        \label{fig:C1C2}
    \normalsize
\end{figure}

\begin{figure}[!h]	
    \footnotesize
    \centering{
        \def\svgwidth{0.95\columnwidth}
        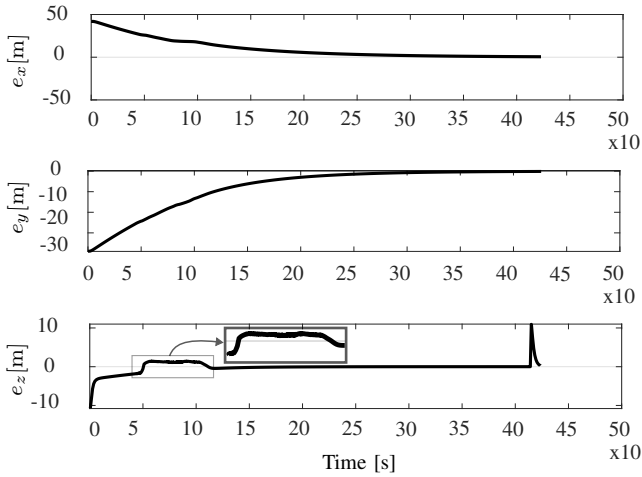}
        \caption{Absolute output tracking error, illustrating the relationship between error reduction and obstacle avoidance. In the figure $e_x = x_I - x_t$, $e_y = y_I - y_t$, and $e_z = z_I - z_t$.
} 
        \label{fig:erro}
    \normalsize
\end{figure}

Fig.~\ref{fig:C1C2} \GR{illustrates that} the thrust forces of the four propellers evolve continuously over time, showing gradual variations during both free flight and avoidance phases\BF{, which occur between $50$ and $120$ seconds}. The absence of high-frequency components in the control \MAS{signal} confirms that the smoothed distance function ensures well-conditioned gradients for the optimization solver. \MAS{The smoothing parameters $\eta$ and $\sigma$ effectively balance geometric accuracy and numerical stability, enabling reliable convergence even under dense sampling and irregular geometries.} This smooth control behavior is significant for real UAV platforms, where actuator bandwidth limitations and mechanical constraints can be \GR{negatively} affected by oscillatory inputs. 

Finally, Fig.~\ref{fig:erro} depicts the tracking error along the mission. The error decreases monotonically after each target transition, stabilizing within the desired tolerance once the UAV reaches the reference position. Even during obstacle avoidance, the transient deviations remain small, evidencing that the smooth \BF{control} barrier \GR{function} formulation effectively balances tracking accuracy and safety. \MAS{For a visual corroboration of this numerical experiment, a illustrative video is available at 
\url{https://youtu.be/zgQIsos1_GI}.}

\begin{figure}[!h]	
    \footnotesize
    \centering{
        \def\svgwidth{0.95\columnwidth}
        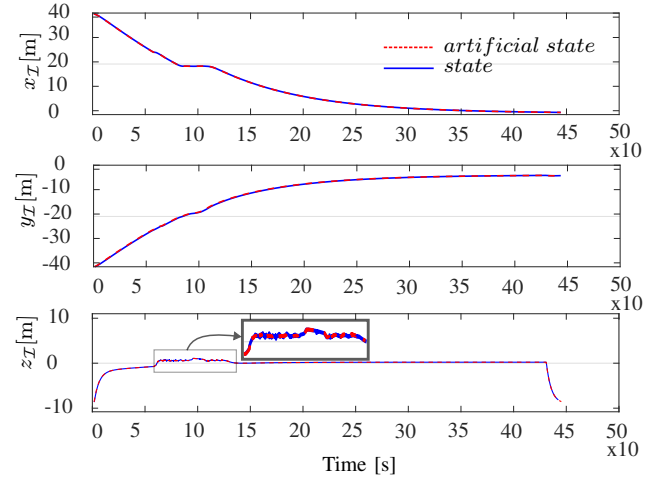}
      \caption{Temporal evolution of the UAV position $x_\mathcal{I}$, $y_\mathcal{I}$, and $z_\mathcal{I}$, when using the classical Euclidean distance in the NMPC formulation.}

        \label{fig:xyzEUC}
    \normalsize
\end{figure}

\begin{figure}[!h]	
    \footnotesize
    \centering{
        \def\svgwidth{0.95\columnwidth}
        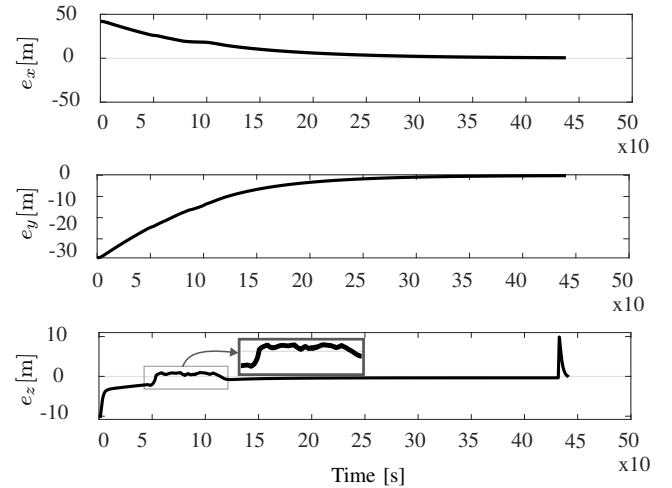}
   \caption{Tracking error evolution during reference transitions, obtained using the classical Euclidean distance.}
        \label{fig:Trackingerro}
    \normalsize
\end{figure}

\begin{figure}[!h]	
    \footnotesize
    \centering{
        \def\svgwidth{0.95\columnwidth}
        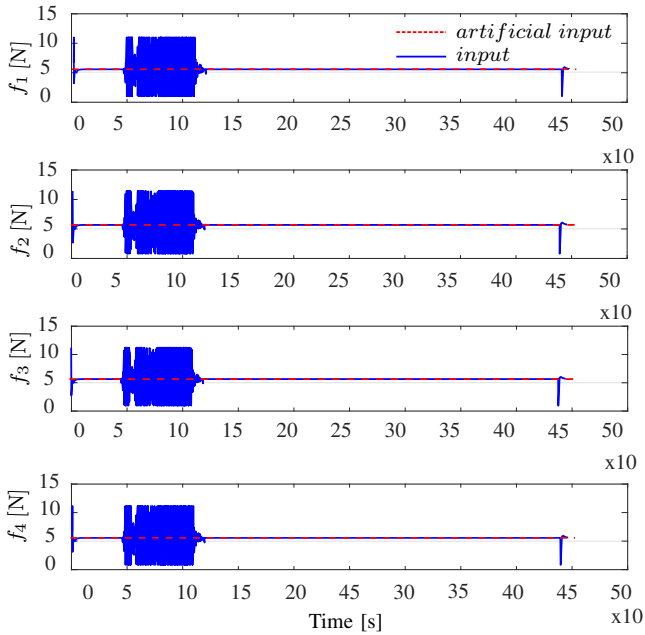}
        \caption{Control inputs applied to the propellers, exhibiting high-frequency variations caused by nondifferentiable gradient transitions inherent to the Euclidean distance formulation.
} 
        \label{fig:U12}
    \normalsize
\end{figure}

\BF{To obtain \GR{comparative} results, \GR{we employed} the same NMPC formulation and parameter settings as in the previous experiment\GR{, but replaced the proposed smoothed metric with the} classical Euclidean distance. Figs.~\ref{fig:xyzEUC} \GR{to} \ref{fig:U12} clearly highlight the advantages of employing the proposed smoothed point-to-cloud distance over the Euclidean metric in the NMPC formulation with CBFs.}

\MAS{When the Euclidean distance is used, the controller exhibits oscillations in the UAV position, particularly near the boundaries of nonfeasible regions, as shown in Fig.~\ref{fig:xyzEUC}. These oscillations arise from the nondifferentiability of the classical Euclidean distance at equidistant, sharp, and concave configurations, where small position changes cause abrupt gradient shifts, introducing numerical stiffness and hindering solver convergence. The tracking error shown in Fig.~\ref{fig:Trackingerro} further highlights these drawbacks, with oscillatory transients appearing near obstacle boundaries. This effect extends to the control inputs, as illustrated in Fig.~\ref{fig:U12}, which exhibit high-frequency variations and chattering behavior, leading to discontinuous avoidance maneuvers and degraded numerical conditioning of the NMPC solution.}

\section{Conclusions}
\label{SecConcl}

This work presented an NMPC strategy for set-point tracking and avoidance of \GR{nonfeasible} regions, integrating smoothed point-to-\GR{cloud} distance metrics with CBFs. The framework was validated in simulation \GR{using} a quadrotor UAV navigating a 3D environment\GR{, where} \MAS{nonfeasible} regions \GR{were} detected online via a virtual LiDAR sensor. The smoothed distance metric allowed the safety constraints to be formulated in a continuously differentiable form\GR{. This enables} their seamless integration as soft penalties in the optimization problem\GR{, which} resulted in well-conditioned gradients, improved numerical stability, and enhanced computational efficiency. \GR{Furthermore, the inclusion of} artificial steady-state variables \GR{ensured the recursive feasibility of the problem} over time, \GR{particularly when tracking} piecewise-constant references. Simulation results confirmed the method’s effectiveness, \GR{showcasing} smooth tracking, consistent constraint satisfaction, and reliable avoidance of complex \GR{nonfeasible} regions. These outcomes demonstrate the potential of the proposed NMPC framework for safe navigation in structured or partially known environments.

Future work will \GR{concentrate} on extending the approach to dynamic scenarios \GR{involving} moving \GR{nonfeasible} regions and \GR{conducting} experimental validation on real robotic platforms. Automatic tuning of smoothing and penalty parameters also represents a promising direction \GR{for improving the framework's} flexibility and \GR{overall} performance.
\bibliographystyle{IEEEtran}
\bibliography{bibtex}

\end{document}

%% file: Figure/t1.pdf_tex
\begingroup%
  \makeatletter%
  \providecommand\color[2][]{%
    \errmessage{(Inkscape) Color is used for the text in Inkscape, but the package 'color.sty' is not loaded}%
    \renewcommand\color[2][]{}%
  }%
  \providecommand\transparent[1]{%
    \errmessage{(Inkscape) Transparency is used (non-zero) for the text in Inkscape, but the package 'transparent.sty' is not loaded}%
    \renewcommand\transparent[1]{}%
  }%
  \providecommand\rotatebox[2]{#2}%
  \newcommand*\fsize{\dimexpr\f@size pt\relax}%
  \newcommand*\lineheight[1]{\fontsize{\fsize}{#1\fsize}\selectfont}%
  \ifx\svgwidth\undefined%
    \setlength{\unitlength}{736.56001282bp}%
    \ifx\svgscale\undefined%
      \relax%
    \else%
      \setlength{\unitlength}{\unitlength * \real{\svgscale}}%
    \fi%
  \else%
    \setlength{\unitlength}{\svgwidth}%
  \fi%
  \global\let\svgwidth\undefined%
  \global\let\svgscale\undefined%
  \makeatother%
  \begin{picture}(1,0.61742422)%
    \lineheight{1}%
    \setlength\tabcolsep{0pt}%
    \put(0,0){\includegraphics[width=\unitlength,page=1]{t1.pdf}}%
    \put(0.30063112,0.04978627){\color[rgb]{0,0,0}\rotatebox{-1.39432488}{\makebox(0,0)[lt]{\lineheight{1.25}\smash{\begin{tabular}[t]{l}$x_\mathcal{I} \rm{[m]}$\end{tabular}}}}}%
    \put(0.8128816,0.09264606){\color[rgb]{0,0,0}\rotatebox{-1.39432488}{\makebox(0,0)[lt]{\lineheight{1.25}\smash{\begin{tabular}[t]{l}$y_\mathcal{I} \rm{[m]}$\end{tabular}}}}}%
    \put(0.02853043,0.29841343){\color[rgb]{0,0,0}\rotatebox{88.60567512}{\makebox(0,0)[lt]{\lineheight{1.25}\smash{\begin{tabular}[t]{l}$z_\mathcal{I} \rm{[m]}$\end{tabular}}}}}%
  \end{picture}%
\endgroup%

%% file: Figure/t2.pdf_tex
\begingroup%
  \makeatletter%
  \providecommand\color[2][]{%
    \errmessage{(Inkscape) Color is used for the text in Inkscape, but the package 'color.sty' is not loaded}%
    \renewcommand\color[2][]{}%
  }%
  \providecommand\transparent[1]{%
    \errmessage{(Inkscape) Transparency is used (non-zero) for the text in Inkscape, but the package 'transparent.sty' is not loaded}%
    \renewcommand\transparent[1]{}%
  }%
  \providecommand\rotatebox[2]{#2}%
  \newcommand*\fsize{\dimexpr\f@size pt\relax}%
  \newcommand*\lineheight[1]{\fontsize{\fsize}{#1\fsize}\selectfont}%
  \ifx\svgwidth\undefined%
    \setlength{\unitlength}{845.68560791bp}%
    \ifx\svgscale\undefined%
      \relax%
    \else%
      \setlength{\unitlength}{\unitlength * \real{\svgscale}}%
    \fi%
  \else%
    \setlength{\unitlength}{\svgwidth}%
  \fi%
  \global\let\svgwidth\undefined%
  \global\let\svgscale\undefined%
  \makeatother%
  \begin{picture}(1,0.5625296)%
    \lineheight{1}%
    \setlength\tabcolsep{0pt}%
    \put(0,0){\includegraphics[width=\unitlength,page=1]{t2.pdf}}%
    \put(0.51641872,0.00485599){\color[rgb]{0,0,0}\rotatebox{-1.39432488}{\makebox(0,0)[lt]{\lineheight{1.25}\smash{\begin{tabular}[t]{l}$x_\mathcal{I} \rm{[m]}$\end{tabular}}}}}%
    \put(0.02193621,0.30108126){\color[rgb]{0,0,0}\rotatebox{88.60567512}{\makebox(0,0)[lt]{\lineheight{1.25}\smash{\begin{tabular}[t]{l}$z_\mathcal{I} \rm{[m]}$\end{tabular}}}}}%
  \end{picture}%
\endgroup%

%% file: Figure/t3.pdf_tex
\begingroup%
  \makeatletter%
  \providecommand\color[2][]{%
    \errmessage{(Inkscape) Color is used for the text in Inkscape, but the package 'color.sty' is not loaded}%
    \renewcommand\color[2][]{}%
  }%
  \providecommand\transparent[1]{%
    \errmessage{(Inkscape) Transparency is used (non-zero) for the text in Inkscape, but the package 'transparent.sty' is not loaded}%
    \renewcommand\transparent[1]{}%
  }%
  \providecommand\rotatebox[2]{#2}%
  \newcommand*\fsize{\dimexpr\f@size pt\relax}%
  \newcommand*\lineheight[1]{\fontsize{\fsize}{#1\fsize}\selectfont}%
  \ifx\svgwidth\undefined%
    \setlength{\unitlength}{526.37997785bp}%
    \ifx\svgscale\undefined%
      \relax%
    \else%
      \setlength{\unitlength}{\unitlength * \real{\svgscale}}%
    \fi%
  \else%
    \setlength{\unitlength}{\svgwidth}%
  \fi%
  \global\let\svgwidth\undefined%
  \global\let\svgscale\undefined%
  \makeatother%
  \begin{picture}(1,0.88339228)%
    \lineheight{1}%
    \setlength\tabcolsep{0pt}%
    \put(0,0){\includegraphics[width=\unitlength,page=1]{t3.pdf}}%
    \put(0.03307077,0.50542002){\color[rgb]{0,0,0}\rotatebox{88.60567512}{\makebox(0,0)[lt]{\lineheight{1.25}\smash{\begin{tabular}[t]{l}$y_\mathcal{I} \rm{[m]}$\end{tabular}}}}}%
    \put(0.51602077,0.01303649){\color[rgb]{0,0,0}\rotatebox{-1.39432488}{\makebox(0,0)[lt]{\lineheight{1.25}\smash{\begin{tabular}[t]{l}$x_\mathcal{I} \rm{[m]}$\end{tabular}}}}}%
  \end{picture}%
\endgroup%

%% file: Figure/C1C2.pdf_tex
\begingroup%
  \makeatletter%
  \providecommand\color[2][]{%
    \errmessage{(Inkscape) Color is used for the text in Inkscape, but the package 'color.sty' is not loaded}%
    \renewcommand\color[2][]{}%
  }%
  \providecommand\transparent[1]{%
    \errmessage{(Inkscape) Transparency is used (non-zero) for the text in Inkscape, but the package 'transparent.sty' is not loaded}%
    \renewcommand\transparent[1]{}%
  }%
  \providecommand\rotatebox[2]{#2}%
  \newcommand*\fsize{\dimexpr\f@size pt\relax}%
  \newcommand*\lineheight[1]{\fontsize{\fsize}{#1\fsize}\selectfont}%
  \ifx\svgwidth\undefined%
    \setlength{\unitlength}{374.02219391bp}%
    \ifx\svgscale\undefined%
      \relax%
    \else%
      \setlength{\unitlength}{\unitlength * \real{\svgscale}}%
    \fi%
  \else%
    \setlength{\unitlength}{\svgwidth}%
  \fi%
  \global\let\svgwidth\undefined%
  \global\let\svgscale\undefined%
  \makeatother%
  \begin{picture}(1,0.88702757)%
    \lineheight{1}%
    \setlength\tabcolsep{0pt}%
    \put(0,0){\includegraphics[width=\unitlength,page=1]{C1C2.pdf}}%
    \put(0.10045483,0.72147961){\color[rgb]{0.14901961,0.14901961,0.14901961}\makebox(0,0)[lt]{\lineheight{1.25}\smash{\begin{tabular}[t]{l}0\end{tabular}}}}%
    \put(0.17216089,0.71771454){\color[rgb]{0.14901961,0.14901961,0.14901961}\makebox(0,0)[lt]{\lineheight{1.25}\smash{\begin{tabular}[t]{l}5\end{tabular}}}}%
    \put(0.25217023,0.71771454){\color[rgb]{0.14901961,0.14901961,0.14901961}\makebox(0,0)[lt]{\lineheight{1.25}\smash{\begin{tabular}[t]{l}10\end{tabular}}}}%
    \put(0.34021259,0.71771454){\color[rgb]{0.14901961,0.14901961,0.14901961}\makebox(0,0)[lt]{\lineheight{1.25}\smash{\begin{tabular}[t]{l}15\end{tabular}}}}%
    \put(0.42906451,0.71771454){\color[rgb]{0.14901961,0.14901961,0.14901961}\makebox(0,0)[lt]{\lineheight{1.25}\smash{\begin{tabular}[t]{l}20\end{tabular}}}}%
    \put(0.51692693,0.71771454){\color[rgb]{0.14901961,0.14901961,0.14901961}\makebox(0,0)[lt]{\lineheight{1.25}\smash{\begin{tabular}[t]{l}25\end{tabular}}}}%
    \put(0.61326003,0.71728651){\color[rgb]{0.14901961,0.14901961,0.14901961}\makebox(0,0)[lt]{\lineheight{1.25}\smash{\begin{tabular}[t]{l}30\end{tabular}}}}%
    \put(0.70141933,0.71788329){\color[rgb]{0.14901961,0.14901961,0.14901961}\makebox(0,0)[lt]{\lineheight{1.25}\smash{\begin{tabular}[t]{l}35\end{tabular}}}}%
    \put(0.78986644,0.71728651){\color[rgb]{0.14901961,0.14901961,0.14901961}\makebox(0,0)[lt]{\lineheight{1.25}\smash{\begin{tabular}[t]{l}40\end{tabular}}}}%
    \put(0.87839458,0.71668976){\color[rgb]{0.14901961,0.14901961,0.14901961}\makebox(0,0)[lt]{\lineheight{1.25}\smash{\begin{tabular}[t]{l}45\end{tabular}}}}%
    \put(0.95741751,0.71728648){\color[rgb]{0.14901961,0.14901961,0.14901961}\makebox(0,0)[lt]{\lineheight{1.25}\smash{\begin{tabular}[t]{l}50\end{tabular}}}}%
    \put(0.95880289,0.68242353){\color[rgb]{0.14901961,0.14901961,0.14901961}\makebox(0,0)[lt]{\lineheight{1.25}\smash{\begin{tabular}[t]{l}x10\end{tabular}}}}%
    \put(0.10045483,0.27000702){\color[rgb]{0.14901961,0.14901961,0.14901961}\makebox(0,0)[lt]{\lineheight{1.25}\smash{\begin{tabular}[t]{l}0\end{tabular}}}}%
    \put(0.17216089,0.27025242){\color[rgb]{0.14901961,0.14901961,0.14901961}\makebox(0,0)[lt]{\lineheight{1.25}\smash{\begin{tabular}[t]{l}5\end{tabular}}}}%
    \put(0.25217018,0.27036161){\color[rgb]{0.14901961,0.14901961,0.14901961}\makebox(0,0)[lt]{\lineheight{1.25}\smash{\begin{tabular}[t]{l}10\end{tabular}}}}%
    \put(0.34021259,0.27025242){\color[rgb]{0.14901961,0.14901961,0.14901961}\makebox(0,0)[lt]{\lineheight{1.25}\smash{\begin{tabular}[t]{l}15\end{tabular}}}}%
    \put(0.42906451,0.27025242){\color[rgb]{0.14901961,0.14901961,0.14901961}\makebox(0,0)[lt]{\lineheight{1.25}\smash{\begin{tabular}[t]{l}20\end{tabular}}}}%
    \put(0.5169269,0.27025242){\color[rgb]{0.14901961,0.14901961,0.14901961}\makebox(0,0)[lt]{\lineheight{1.25}\smash{\begin{tabular}[t]{l}25\end{tabular}}}}%
    \put(0.60523918,0.2698244){\color[rgb]{0.14901961,0.14901961,0.14901961}\makebox(0,0)[lt]{\lineheight{1.25}\smash{\begin{tabular}[t]{l}30\end{tabular}}}}%
    \put(0.69339848,0.27042117){\color[rgb]{0.14901961,0.14901961,0.14901961}\makebox(0,0)[lt]{\lineheight{1.25}\smash{\begin{tabular}[t]{l}35\end{tabular}}}}%
    \put(0.78184559,0.2698244){\color[rgb]{0.14901961,0.14901961,0.14901961}\makebox(0,0)[lt]{\lineheight{1.25}\smash{\begin{tabular}[t]{l}40\end{tabular}}}}%
    \put(0.87037367,0.27323816){\color[rgb]{0.14901961,0.14901961,0.14901961}\makebox(0,0)[lt]{\lineheight{1.25}\smash{\begin{tabular}[t]{l}45\end{tabular}}}}%
    \put(0.95741751,0.27383481){\color[rgb]{0.14901961,0.14901961,0.14901961}\makebox(0,0)[lt]{\lineheight{1.25}\smash{\begin{tabular}[t]{l}50\end{tabular}}}}%
    \put(0.95843379,0.22049685){\color[rgb]{0.14901961,0.14901961,0.14901961}\makebox(0,0)[lt]{\lineheight{1.25}\smash{\begin{tabular}[t]{l}x10\end{tabular}}}}%
    \put(0.09641134,0.49508549){\color[rgb]{0.14901961,0.14901961,0.14901961}\makebox(0,0)[lt]{\lineheight{1.25}\smash{\begin{tabular}[t]{l}0\end{tabular}}}}%
    \put(0.17216089,0.49533089){\color[rgb]{0.14901961,0.14901961,0.14901961}\makebox(0,0)[lt]{\lineheight{1.25}\smash{\begin{tabular}[t]{l}5\end{tabular}}}}%
    \put(0.24812669,0.49533089){\color[rgb]{0.14901961,0.14901961,0.14901961}\makebox(0,0)[lt]{\lineheight{1.25}\smash{\begin{tabular}[t]{l}10\end{tabular}}}}%
    \put(0.33616907,0.49533089){\color[rgb]{0.14901961,0.14901961,0.14901961}\makebox(0,0)[lt]{\lineheight{1.25}\smash{\begin{tabular}[t]{l}15\end{tabular}}}}%
    \put(0.42502103,0.49533089){\color[rgb]{0.14901961,0.14901961,0.14901961}\makebox(0,0)[lt]{\lineheight{1.25}\smash{\begin{tabular}[t]{l}20\end{tabular}}}}%
    \put(0.51288341,0.49533089){\color[rgb]{0.14901961,0.14901961,0.14901961}\makebox(0,0)[lt]{\lineheight{1.25}\smash{\begin{tabular}[t]{l}25\end{tabular}}}}%
    \put(0.60119569,0.49490285){\color[rgb]{0.14901961,0.14901961,0.14901961}\makebox(0,0)[lt]{\lineheight{1.25}\smash{\begin{tabular}[t]{l}30\end{tabular}}}}%
    \put(0.68935493,0.49549964){\color[rgb]{0.14901961,0.14901961,0.14901961}\makebox(0,0)[lt]{\lineheight{1.25}\smash{\begin{tabular}[t]{l}35\end{tabular}}}}%
    \put(0.77780211,0.49490285){\color[rgb]{0.14901961,0.14901961,0.14901961}\makebox(0,0)[lt]{\lineheight{1.25}\smash{\begin{tabular}[t]{l}40\end{tabular}}}}%
    \put(0.86633018,0.49430611){\color[rgb]{0.14901961,0.14901961,0.14901961}\makebox(0,0)[lt]{\lineheight{1.25}\smash{\begin{tabular}[t]{l}45\end{tabular}}}}%
    \put(0.95337402,0.49490282){\color[rgb]{0.14901961,0.14901961,0.14901961}\makebox(0,0)[lt]{\lineheight{1.25}\smash{\begin{tabular}[t]{l}50\end{tabular}}}}%
    \put(0.9507493,0.45156576){\color[rgb]{0.14901961,0.14901961,0.14901961}\makebox(0,0)[lt]{\lineheight{1.25}\smash{\begin{tabular}[t]{l}x10\end{tabular}}}}%
    \put(0.10652007,0.048328){\color[rgb]{0.14901961,0.14901961,0.14901961}\makebox(0,0)[lt]{\lineheight{1.25}\smash{\begin{tabular}[t]{l}0\end{tabular}}}}%
    \put(0.17418263,0.04857347){\color[rgb]{0.14901961,0.14901961,0.14901961}\makebox(0,0)[lt]{\lineheight{1.25}\smash{\begin{tabular}[t]{l}5\end{tabular}}}}%
    \put(0.26632244,0.04857347){\color[rgb]{0.14901961,0.14901961,0.14901961}\makebox(0,0)[lt]{\lineheight{1.25}\smash{\begin{tabular}[t]{l}10\end{tabular}}}}%
    \put(0.35436479,0.04857347){\color[rgb]{0.14901961,0.14901961,0.14901961}\makebox(0,0)[lt]{\lineheight{1.25}\smash{\begin{tabular}[t]{l}15\end{tabular}}}}%
    \put(0.44321671,0.04857347){\color[rgb]{0.14901961,0.14901961,0.14901961}\makebox(0,0)[lt]{\lineheight{1.25}\smash{\begin{tabular}[t]{l}20\end{tabular}}}}%
    \put(0.53107913,0.04857347){\color[rgb]{0.14901961,0.14901961,0.14901961}\makebox(0,0)[lt]{\lineheight{1.25}\smash{\begin{tabular}[t]{l}25\end{tabular}}}}%
    \put(0.61939138,0.04814539){\color[rgb]{0.14901961,0.14901961,0.14901961}\makebox(0,0)[lt]{\lineheight{1.25}\smash{\begin{tabular}[t]{l}30\end{tabular}}}}%
    \put(0.70755068,0.04874221){\color[rgb]{0.14901961,0.14901961,0.14901961}\makebox(0,0)[lt]{\lineheight{1.25}\smash{\begin{tabular}[t]{l}35\end{tabular}}}}%
    \put(0.79599786,0.04814539){\color[rgb]{0.14901961,0.14901961,0.14901961}\makebox(0,0)[lt]{\lineheight{1.25}\smash{\begin{tabular}[t]{l}40\end{tabular}}}}%
    \put(0.88452593,0.04754868){\color[rgb]{0.14901961,0.14901961,0.14901961}\makebox(0,0)[lt]{\lineheight{1.25}\smash{\begin{tabular}[t]{l}45\end{tabular}}}}%
    \put(0.95943925,0.04814533){\color[rgb]{0.14901961,0.14901961,0.14901961}\makebox(0,0)[lt]{\lineheight{1.25}\smash{\begin{tabular}[t]{l}50\end{tabular}}}}%
    \put(0.96082463,0.01161767){\color[rgb]{0.14901961,0.14901961,0.14901961}\makebox(0,0)[lt]{\lineheight{1.25}\smash{\begin{tabular}[t]{l}x10\end{tabular}}}}%
    \put(0.04276547,0.38736268){\color[rgb]{0.14901961,0.14901961,0.14901961}\makebox(0,0)[lt]{\lineheight{1.25}\smash{\begin{tabular}[t]{l}10\end{tabular}}}}%
    \put(0.04188985,0.42528052){\color[rgb]{0.14901961,0.14901961,0.14901961}\makebox(0,0)[lt]{\lineheight{1.25}\smash{\begin{tabular}[t]{l}15\end{tabular}}}}%
    \put(0.05263665,0.34968533){\color[rgb]{0.14901961,0.14901961,0.14901961}\makebox(0,0)[lt]{\lineheight{1.25}\smash{\begin{tabular}[t]{l}5\end{tabular}}}}%
    \put(0.05263665,0.31176749){\color[rgb]{0.14901961,0.14901961,0.14901961}\makebox(0,0)[lt]{\lineheight{1.25}\smash{\begin{tabular}[t]{l}0\end{tabular}}}}%
    \put(0.0411571,0.83713226){\color[rgb]{0.14901961,0.14901961,0.14901961}\makebox(0,0)[lt]{\lineheight{1.25}\smash{\begin{tabular}[t]{l}10\end{tabular}}}}%
    \put(0.04028147,0.87505013){\color[rgb]{0.14901961,0.14901961,0.14901961}\makebox(0,0)[lt]{\lineheight{1.25}\smash{\begin{tabular}[t]{l}15\end{tabular}}}}%
    \put(0.04698478,0.7994549){\color[rgb]{0.14901961,0.14901961,0.14901961}\makebox(0,0)[lt]{\lineheight{1.25}\smash{\begin{tabular}[t]{l}5\end{tabular}}}}%
    \put(0.04698478,0.76153703){\color[rgb]{0.14901961,0.14901961,0.14901961}\makebox(0,0)[lt]{\lineheight{1.25}\smash{\begin{tabular}[t]{l}0\end{tabular}}}}%
    \put(0.04179594,0.61079374){\color[rgb]{0.14901961,0.14901961,0.14901961}\makebox(0,0)[lt]{\lineheight{1.25}\smash{\begin{tabular}[t]{l}10\end{tabular}}}}%
    \put(0.04092032,0.64871161){\color[rgb]{0.14901961,0.14901961,0.14901961}\makebox(0,0)[lt]{\lineheight{1.25}\smash{\begin{tabular}[t]{l}15\end{tabular}}}}%
    \put(0.05166711,0.57311637){\color[rgb]{0.14901961,0.14901961,0.14901961}\makebox(0,0)[lt]{\lineheight{1.25}\smash{\begin{tabular}[t]{l}5\end{tabular}}}}%
    \put(0.05166711,0.53519853){\color[rgb]{0.14901961,0.14901961,0.14901961}\makebox(0,0)[lt]{\lineheight{1.25}\smash{\begin{tabular}[t]{l}0\end{tabular}}}}%
    \put(0.04517805,0.16449637){\color[rgb]{0.14901961,0.14901961,0.14901961}\makebox(0,0)[lt]{\lineheight{1.25}\smash{\begin{tabular}[t]{l}10\end{tabular}}}}%
    \put(0.04430242,0.20241427){\color[rgb]{0.14901961,0.14901961,0.14901961}\makebox(0,0)[lt]{\lineheight{1.25}\smash{\begin{tabular}[t]{l}15\end{tabular}}}}%
    \put(0.05504921,0.12681903){\color[rgb]{0.14901961,0.14901961,0.14901961}\makebox(0,0)[lt]{\lineheight{1.25}\smash{\begin{tabular}[t]{l}5\end{tabular}}}}%
    \put(0.05504921,0.08890113){\color[rgb]{0.14901961,0.14901961,0.14901961}\makebox(0,0)[lt]{\lineheight{1.25}\smash{\begin{tabular}[t]{l}0\end{tabular}}}}%
    \put(0,0){\includegraphics[width=\unitlength,page=2]{C1C2.pdf}}%
    \put(0.69635846,0.8460099){\color[rgb]{0,0,0}\makebox(0,0)[lt]{\lineheight{1.25}\smash{\begin{tabular}[t]{l}$artificial\;input$\end{tabular}}}}%
    \put(0.69635846,0.81410913){\color[rgb]{0,0,0}\makebox(0,0)[lt]{\lineheight{1.25}\smash{\begin{tabular}[t]{l}$input$\end{tabular}}}}%
    \put(0.48193079,0.0068227){\color[rgb]{0,0,0}\makebox(0,0)[lt]{\lineheight{1.25}\smash{\begin{tabular}[t]{l}Time [s]\end{tabular}}}}%
    \put(0.01795414,0.56602106){\color[rgb]{0,0,0}\rotatebox{87.62495898}{\makebox(0,0)[lt]{\lineheight{1.25}\smash{\begin{tabular}[t]{l}${f}_2 \, \si{[\newton]}$\end{tabular}}}}}%
    \put(0.01943778,0.78446266){\color[rgb]{0,0,0}\rotatebox{86.06003741}{\makebox(0,0)[lt]{\lineheight{1.25}\smash{\begin{tabular}[t]{l}${f}_1 \, \si{[\newton]}$\end{tabular}}}}}%
    \put(0.02213429,0.38922534){\color[rgb]{0,0,0}\rotatebox{89.40353837}{\makebox(0,0)[lt]{\lineheight{1.25}\smash{\begin{tabular}[t]{l}${f}_3 \, \si{[\newton]}$\end{tabular}}}}}%
    \put(0.0208241,0.16905515){\color[rgb]{0,0,0}\rotatebox{87.9706991}{\makebox(0,0)[lt]{\lineheight{1.25}\smash{\begin{tabular}[t]{l}${f}_4 \, \si{[\newton]}$\end{tabular}}}}}%
    \put(0,0){\includegraphics[width=\unitlength,page=3]{C1C2.pdf}}%
  \end{picture}%
\endgroup%

%% file: Figure/errotargetSmoothed.pdf_tex
\begingroup%
  \makeatletter%
  \providecommand\color[2][]{%
    \errmessage{(Inkscape) Color is used for the text in Inkscape, but the package 'color.sty' is not loaded}%
    \renewcommand\color[2][]{}%
  }%
  \providecommand\transparent[1]{%
    \errmessage{(Inkscape) Transparency is used (non-zero) for the text in Inkscape, but the package 'transparent.sty' is not loaded}%
    \renewcommand\transparent[1]{}%
  }%
  \providecommand\rotatebox[2]{#2}%
  \newcommand*\fsize{\dimexpr\f@size pt\relax}%
  \newcommand*\lineheight[1]{\fontsize{\fsize}{#1\fsize}\selectfont}%
  \ifx\svgwidth\undefined%
    \setlength{\unitlength}{379.59597264bp}%
    \ifx\svgscale\undefined%
      \relax%
    \else%
      \setlength{\unitlength}{\unitlength * \real{\svgscale}}%
    \fi%
  \else%
    \setlength{\unitlength}{\svgwidth}%
  \fi%
  \global\let\svgwidth\undefined%
  \global\let\svgscale\undefined%
  \makeatother%
  \begin{picture}(1,0.743501)%
    \lineheight{1}%
    \setlength\tabcolsep{0pt}%
    \put(0.03482439,0.10512807){\color[rgb]{0.14901961,0.14901961,0.14901961}\makebox(0,0)[lt]{\lineheight{1.25}\smash{\begin{tabular}[t]{l}-10\end{tabular}}}}%
    \put(0.02100907,0.62175273){\color[rgb]{0,0,0}\rotatebox{88.26183689}{\makebox(0,0)[lt]{\lineheight{1.25}\smash{\begin{tabular}[t]{l}$e_x \rm{[m]}$\end{tabular}}}}}%
    \put(0.02100904,0.38596202){\color[rgb]{0,0,0}\rotatebox{88.26176777}{\makebox(0,0)[lt]{\lineheight{1.25}\smash{\begin{tabular}[t]{l}$e_y \rm{[m]}$\end{tabular}}}}}%
    \put(0.02038725,0.12515026){\color[rgb]{0,0,0}\rotatebox{88.846491}{\makebox(0,0)[lt]{\lineheight{1.25}\smash{\begin{tabular}[t]{l}$e_z \rm{[m]}$\end{tabular}}}}}%
    \put(0.5047482,0.00691062){\color[rgb]{0,0,0}\makebox(0,0)[lt]{\lineheight{1.25}\smash{\begin{tabular}[t]{l}Time [s]\end{tabular}}}}%
    \put(0.05357295,0.59615089){\color[rgb]{0.14901961,0.14901961,0.14901961}\makebox(0,0)[lt]{\lineheight{1.25}\smash{\begin{tabular}[t]{l}-50\end{tabular}}}}%
    \put(0,0){\includegraphics[width=\unitlength,page=1]{errotargetSmoothed.pdf}}%
    \put(0.12722162,0.56082915){\color[rgb]{0.14901961,0.14901961,0.14901961}\makebox(0,0)[lt]{\lineheight{1.25}\smash{\begin{tabular}[t]{l}0\end{tabular}}}}%
    \put(0.18939445,0.56107769){\color[rgb]{0.14901961,0.14901961,0.14901961}\makebox(0,0)[lt]{\lineheight{1.25}\smash{\begin{tabular}[t]{l}5\end{tabular}}}}%
    \put(0.27548805,0.56107769){\color[rgb]{0.14901961,0.14901961,0.14901961}\makebox(0,0)[lt]{\lineheight{1.25}\smash{\begin{tabular}[t]{l}10\end{tabular}}}}%
    \put(0.36152893,0.56107769){\color[rgb]{0.14901961,0.14901961,0.14901961}\makebox(0,0)[lt]{\lineheight{1.25}\smash{\begin{tabular}[t]{l}15\end{tabular}}}}%
    \put(0.44836095,0.56107769){\color[rgb]{0.14901961,0.14901961,0.14901961}\makebox(0,0)[lt]{\lineheight{1.25}\smash{\begin{tabular}[t]{l}20\end{tabular}}}}%
    \put(0.534226,0.56107769){\color[rgb]{0.14901961,0.14901961,0.14901961}\makebox(0,0)[lt]{\lineheight{1.25}\smash{\begin{tabular}[t]{l}25\end{tabular}}}}%
    \put(0.62053065,0.56064414){\color[rgb]{0.14901961,0.14901961,0.14901961}\makebox(0,0)[lt]{\lineheight{1.25}\smash{\begin{tabular}[t]{l}30\end{tabular}}}}%
    \put(0.70668586,0.56124862){\color[rgb]{0.14901961,0.14901961,0.14901961}\makebox(0,0)[lt]{\lineheight{1.25}\smash{\begin{tabular}[t]{l}35\end{tabular}}}}%
    \put(0.79312238,0.56064414){\color[rgb]{0.14901961,0.14901961,0.14901961}\makebox(0,0)[lt]{\lineheight{1.25}\smash{\begin{tabular}[t]{l}40\end{tabular}}}}%
    \put(0.87963802,0.56003969){\color[rgb]{0.14901961,0.14901961,0.14901961}\makebox(0,0)[lt]{\lineheight{1.25}\smash{\begin{tabular}[t]{l}45\end{tabular}}}}%
    \put(0.96470304,0.56064411){\color[rgb]{0.14901961,0.14901961,0.14901961}\makebox(0,0)[lt]{\lineheight{1.25}\smash{\begin{tabular}[t]{l}50\end{tabular}}}}%
    \put(0.96566393,0.52319894){\color[rgb]{0.14901961,0.14901961,0.14901961}\makebox(0,0)[lt]{\lineheight{1.25}\smash{\begin{tabular}[t]{l}x10\end{tabular}}}}%
    \put(0.11931856,0.31188024){\color[rgb]{0.14901961,0.14901961,0.14901961}\makebox(0,0)[lt]{\lineheight{1.25}\smash{\begin{tabular}[t]{l}0\end{tabular}}}}%
    \put(0.18149131,0.31212879){\color[rgb]{0.14901961,0.14901961,0.14901961}\makebox(0,0)[lt]{\lineheight{1.25}\smash{\begin{tabular}[t]{l}5\end{tabular}}}}%
    \put(0.26758485,0.31212879){\color[rgb]{0.14901961,0.14901961,0.14901961}\makebox(0,0)[lt]{\lineheight{1.25}\smash{\begin{tabular}[t]{l}10\end{tabular}}}}%
    \put(0.35362562,0.31212879){\color[rgb]{0.14901961,0.14901961,0.14901961}\makebox(0,0)[lt]{\lineheight{1.25}\smash{\begin{tabular}[t]{l}15\end{tabular}}}}%
    \put(0.44045758,0.31212879){\color[rgb]{0.14901961,0.14901961,0.14901961}\makebox(0,0)[lt]{\lineheight{1.25}\smash{\begin{tabular}[t]{l}20\end{tabular}}}}%
    \put(0.52632254,0.31212879){\color[rgb]{0.14901961,0.14901961,0.14901961}\makebox(0,0)[lt]{\lineheight{1.25}\smash{\begin{tabular}[t]{l}25\end{tabular}}}}%
    \put(0.61262719,0.31169526){\color[rgb]{0.14901961,0.14901961,0.14901961}\makebox(0,0)[lt]{\lineheight{1.25}\smash{\begin{tabular}[t]{l}30\end{tabular}}}}%
    \put(0.6987824,0.31229974){\color[rgb]{0.14901961,0.14901961,0.14901961}\makebox(0,0)[lt]{\lineheight{1.25}\smash{\begin{tabular}[t]{l}35\end{tabular}}}}%
    \put(0.78521899,0.31169526){\color[rgb]{0.14901961,0.14901961,0.14901961}\makebox(0,0)[lt]{\lineheight{1.25}\smash{\begin{tabular}[t]{l}40\end{tabular}}}}%
    \put(0.87173462,0.31109079){\color[rgb]{0.14901961,0.14901961,0.14901961}\makebox(0,0)[lt]{\lineheight{1.25}\smash{\begin{tabular}[t]{l}45\end{tabular}}}}%
    \put(0.95679965,0.3116952){\color[rgb]{0.14901961,0.14901961,0.14901961}\makebox(0,0)[lt]{\lineheight{1.25}\smash{\begin{tabular}[t]{l}50\end{tabular}}}}%
    \put(0.95776053,0.27425002){\color[rgb]{0.14901961,0.14901961,0.14901961}\makebox(0,0)[lt]{\lineheight{1.25}\smash{\begin{tabular}[t]{l}x10\end{tabular}}}}%
    \put(0.12919742,0.06095555){\color[rgb]{0.14901961,0.14901961,0.14901961}\makebox(0,0)[lt]{\lineheight{1.25}\smash{\begin{tabular}[t]{l}0\end{tabular}}}}%
    \put(0.19137024,0.0612041){\color[rgb]{0.14901961,0.14901961,0.14901961}\makebox(0,0)[lt]{\lineheight{1.25}\smash{\begin{tabular}[t]{l}5\end{tabular}}}}%
    \put(0.27746384,0.0612041){\color[rgb]{0.14901961,0.14901961,0.14901961}\makebox(0,0)[lt]{\lineheight{1.25}\smash{\begin{tabular}[t]{l}10\end{tabular}}}}%
    \put(0.36350471,0.0612041){\color[rgb]{0.14901961,0.14901961,0.14901961}\makebox(0,0)[lt]{\lineheight{1.25}\smash{\begin{tabular}[t]{l}15\end{tabular}}}}%
    \put(0.45033673,0.0612041){\color[rgb]{0.14901961,0.14901961,0.14901961}\makebox(0,0)[lt]{\lineheight{1.25}\smash{\begin{tabular}[t]{l}20\end{tabular}}}}%
    \put(0.53620172,0.0612041){\color[rgb]{0.14901961,0.14901961,0.14901961}\makebox(0,0)[lt]{\lineheight{1.25}\smash{\begin{tabular}[t]{l}25\end{tabular}}}}%
    \put(0.62250637,0.06077058){\color[rgb]{0.14901961,0.14901961,0.14901961}\makebox(0,0)[lt]{\lineheight{1.25}\smash{\begin{tabular}[t]{l}30\end{tabular}}}}%
    \put(0.70866158,0.06137505){\color[rgb]{0.14901961,0.14901961,0.14901961}\makebox(0,0)[lt]{\lineheight{1.25}\smash{\begin{tabular}[t]{l}35\end{tabular}}}}%
    \put(0.79509811,0.06077058){\color[rgb]{0.14901961,0.14901961,0.14901961}\makebox(0,0)[lt]{\lineheight{1.25}\smash{\begin{tabular}[t]{l}40\end{tabular}}}}%
    \put(0.88161374,0.0601661){\color[rgb]{0.14901961,0.14901961,0.14901961}\makebox(0,0)[lt]{\lineheight{1.25}\smash{\begin{tabular}[t]{l}45\end{tabular}}}}%
    \put(0.96667877,0.06077052){\color[rgb]{0.14901961,0.14901961,0.14901961}\makebox(0,0)[lt]{\lineheight{1.25}\smash{\begin{tabular}[t]{l}50\end{tabular}}}}%
    \put(0.96763965,0.02332542){\color[rgb]{0.14901961,0.14901961,0.14901961}\makebox(0,0)[lt]{\lineheight{1.25}\smash{\begin{tabular}[t]{l}x10\end{tabular}}}}%
    \put(0.0472996,0.43966543){\color[rgb]{0.14901961,0.14901961,0.14901961}\makebox(0,0)[lt]{\lineheight{1.25}\smash{\begin{tabular}[t]{l}-10\end{tabular}}}}%
    \put(0.06620173,0.48202359){\color[rgb]{0.14901961,0.14901961,0.14901961}\makebox(0,0)[lt]{\lineheight{1.25}\smash{\begin{tabular}[t]{l}0\end{tabular}}}}%
    \put(0.04904322,0.40150242){\color[rgb]{0.14901961,0.14901961,0.14901961}\makebox(0,0)[lt]{\lineheight{1.25}\smash{\begin{tabular}[t]{l}-20\end{tabular}}}}%
    \put(0.04904322,0.35914427){\color[rgb]{0.14901961,0.14901961,0.14901961}\makebox(0,0)[lt]{\lineheight{1.25}\smash{\begin{tabular}[t]{l}-30\end{tabular}}}}%
    \put(0.04409801,0.22734093){\color[rgb]{0.14901961,0.14901961,0.14901961}\makebox(0,0)[lt]{\lineheight{1.25}\smash{\begin{tabular}[t]{l}10\end{tabular}}}}%
    \put(0.05609338,0.16411591){\color[rgb]{0.14901961,0.14901961,0.14901961}\makebox(0,0)[lt]{\lineheight{1.25}\smash{\begin{tabular}[t]{l}0\end{tabular}}}}%
    \put(0.05990428,0.73116613){\color[rgb]{0.14901961,0.14901961,0.14901961}\makebox(0,0)[lt]{\lineheight{1.25}\smash{\begin{tabular}[t]{l}50\end{tabular}}}}%
    \put(0.07189965,0.66596519){\color[rgb]{0.14901961,0.14901961,0.14901961}\makebox(0,0)[lt]{\lineheight{1.25}\smash{\begin{tabular}[t]{l}0\end{tabular}}}}%
    \put(0,0){\includegraphics[width=\unitlength,page=2]{errotargetSmoothed.pdf}}%
  \end{picture}%
\endgroup%

%% file: Figure/xyzEUC.pdf_tex
\begingroup%
  \makeatletter%
  \providecommand\color[2][]{%
    \errmessage{(Inkscape) Color is used for the text in Inkscape, but the package 'color.sty' is not loaded}%
    \renewcommand\color[2][]{}%
  }%
  \providecommand\transparent[1]{%
    \errmessage{(Inkscape) Transparency is used (non-zero) for the text in Inkscape, but the package 'transparent.sty' is not loaded}%
    \renewcommand\transparent[1]{}%
  }%
  \providecommand\rotatebox[2]{#2}%
  \newcommand*\fsize{\dimexpr\f@size pt\relax}%
  \newcommand*\lineheight[1]{\fontsize{\fsize}{#1\fsize}\selectfont}%
  \ifx\svgwidth\undefined%
    \setlength{\unitlength}{375.89250183bp}%
    \ifx\svgscale\undefined%
      \relax%
    \else%
      \setlength{\unitlength}{\unitlength * \real{\svgscale}}%
    \fi%
  \else%
    \setlength{\unitlength}{\svgwidth}%
  \fi%
  \global\let\svgwidth\undefined%
  \global\let\svgscale\undefined%
  \makeatother%
  \begin{picture}(1,0.76739357)%
    \lineheight{1}%
    \setlength\tabcolsep{0pt}%
    \put(0,0){\includegraphics[width=\unitlength,page=1]{xyzEUC.pdf}}%
    \put(0.49036625,0.01089539){\color[rgb]{0,0,0}\makebox(0,0)[lt]{\lineheight{1.25}\smash{\begin{tabular}[t]{l}Time [s]\end{tabular}}}}%
    \put(0.02574903,0.62617118){\color[rgb]{0,0,0}\rotatebox{88.60567512}{\makebox(0,0)[lt]{\lineheight{1.25}\smash{\begin{tabular}[t]{l}$x_\mathcal{I} \rm{[m]}$\end{tabular}}}}}%
    \put(0.01932813,0.40259275){\color[rgb]{0,0,0}\rotatebox{88.60567512}{\makebox(0,0)[lt]{\lineheight{1.25}\smash{\begin{tabular}[t]{l}$y_\mathcal{I} \rm{[m]}$\end{tabular}}}}}%
    \put(0,0){\includegraphics[width=\unitlength,page=2]{xyzEUC.pdf}}%
    \put(0.68578277,0.68650732){\color[rgb]{0,0,0}\makebox(0,0)[lt]{\lineheight{1.25}\smash{\begin{tabular}[t]{l}$artificial\;state$\end{tabular}}}}%
    \put(0.68578277,0.65476528){\color[rgb]{0,0,0}\makebox(0,0)[lt]{\lineheight{1.25}\smash{\begin{tabular}[t]{l}$state$\end{tabular}}}}%
    \put(0,0){\includegraphics[width=\unitlength,page=3]{xyzEUC.pdf}}%
    \put(0.1185641,0.06506212){\color[rgb]{0.14901961,0.14901961,0.14901961}\makebox(0,0)[lt]{\lineheight{1.25}\smash{\begin{tabular}[t]{l}0\end{tabular}}}}%
    \put(0.18011632,0.06535071){\color[rgb]{0.14901961,0.14901961,0.14901961}\makebox(0,0)[lt]{\lineheight{1.25}\smash{\begin{tabular}[t]{l}5\end{tabular}}}}%
    \put(0.2653506,0.06535071){\color[rgb]{0.14901961,0.14901961,0.14901961}\makebox(0,0)[lt]{\lineheight{1.25}\smash{\begin{tabular}[t]{l}10\end{tabular}}}}%
    \put(0.35053266,0.06535071){\color[rgb]{0.14901961,0.14901961,0.14901961}\makebox(0,0)[lt]{\lineheight{1.25}\smash{\begin{tabular}[t]{l}15\end{tabular}}}}%
    \put(0.43649796,0.06535071){\color[rgb]{0.14901961,0.14901961,0.14901961}\makebox(0,0)[lt]{\lineheight{1.25}\smash{\begin{tabular}[t]{l}20\end{tabular}}}}%
    \put(0.5215059,0.06535071){\color[rgb]{0.14901961,0.14901961,0.14901961}\makebox(0,0)[lt]{\lineheight{1.25}\smash{\begin{tabular}[t]{l}25\end{tabular}}}}%
    \put(0.60694899,0.06484736){\color[rgb]{0.14901961,0.14901961,0.14901961}\makebox(0,0)[lt]{\lineheight{1.25}\smash{\begin{tabular}[t]{l}30\end{tabular}}}}%
    \put(0.69224435,0.06554912){\color[rgb]{0.14901961,0.14901961,0.14901961}\makebox(0,0)[lt]{\lineheight{1.25}\smash{\begin{tabular}[t]{l}35\end{tabular}}}}%
    \put(0.77781803,0.06484736){\color[rgb]{0.14901961,0.14901961,0.14901961}\makebox(0,0)[lt]{\lineheight{1.25}\smash{\begin{tabular}[t]{l}40\end{tabular}}}}%
    \put(0.86347005,0.06414566){\color[rgb]{0.14901961,0.14901961,0.14901961}\makebox(0,0)[lt]{\lineheight{1.25}\smash{\begin{tabular}[t]{l}45\end{tabular}}}}%
    \put(0.94768607,0.0648473){\color[rgb]{0.14901961,0.14901961,0.14901961}\makebox(0,0)[lt]{\lineheight{1.25}\smash{\begin{tabular}[t]{l}50\end{tabular}}}}%
    \put(0.94863734,0.03513964){\color[rgb]{0.14901961,0.14901961,0.14901961}\makebox(0,0)[lt]{\lineheight{1.25}\smash{\begin{tabular}[t]{l}x10\end{tabular}}}}%
    \put(0.02036525,0.17423087){\color[rgb]{0,0,0}\rotatebox{88.60567512}{\makebox(0,0)[lt]{\lineheight{1.25}\smash{\begin{tabular}[t]{l}$z_\mathcal{I} \rm{[m]}$\end{tabular}}}}}%
    \put(0.04162749,0.10461245){\color[rgb]{0.14901961,0.14901961,0.14901961}\makebox(0,0)[lt]{\lineheight{1.25}\smash{\begin{tabular}[t]{l}-10\end{tabular}}}}%
    \put(0.04557281,0.17569524){\color[rgb]{0.14901961,0.14901961,0.14901961}\makebox(0,0)[lt]{\lineheight{1.25}\smash{\begin{tabular}[t]{l}0\end{tabular}}}}%
    \put(0.04717805,0.24907065){\color[rgb]{0.14901961,0.14901961,0.14901961}\makebox(0,0)[lt]{\lineheight{1.25}\smash{\begin{tabular}[t]{l}10\end{tabular}}}}%
    \put(0,0){\includegraphics[width=\unitlength,page=4]{xyzEUC.pdf}}%
    \put(0.87012387,0.30137274){\color[rgb]{0.14901961,0.14901961,0.14901961}\makebox(0,0)[lt]{\lineheight{1.25}\smash{\begin{tabular}[t]{l}45\end{tabular}}}}%
    \put(0.95433986,0.30209956){\color[rgb]{0.14901961,0.14901961,0.14901961}\makebox(0,0)[lt]{\lineheight{1.25}\smash{\begin{tabular}[t]{l}50\end{tabular}}}}%
    \put(0.95529117,0.27132531){\color[rgb]{0.14901961,0.14901961,0.14901961}\makebox(0,0)[lt]{\lineheight{1.25}\smash{\begin{tabular}[t]{l}x10\end{tabular}}}}%
    \put(0,0){\includegraphics[width=\unitlength,page=5]{xyzEUC.pdf}}%
    \put(0.12051098,0.29741519){\color[rgb]{0.14901961,0.14901961,0.14901961}\makebox(0,0)[lt]{\lineheight{1.25}\smash{\begin{tabular}[t]{l}0\end{tabular}}}}%
    \put(0.1820632,0.29771413){\color[rgb]{0.14901961,0.14901961,0.14901961}\makebox(0,0)[lt]{\lineheight{1.25}\smash{\begin{tabular}[t]{l}5\end{tabular}}}}%
    \put(0.26729749,0.29771413){\color[rgb]{0.14901961,0.14901961,0.14901961}\makebox(0,0)[lt]{\lineheight{1.25}\smash{\begin{tabular}[t]{l}10\end{tabular}}}}%
    \put(0.35247956,0.29771413){\color[rgb]{0.14901961,0.14901961,0.14901961}\makebox(0,0)[lt]{\lineheight{1.25}\smash{\begin{tabular}[t]{l}15\end{tabular}}}}%
    \put(0.43844487,0.29771413){\color[rgb]{0.14901961,0.14901961,0.14901961}\makebox(0,0)[lt]{\lineheight{1.25}\smash{\begin{tabular}[t]{l}20\end{tabular}}}}%
    \put(0.52345271,0.29771413){\color[rgb]{0.14901961,0.14901961,0.14901961}\makebox(0,0)[lt]{\lineheight{1.25}\smash{\begin{tabular}[t]{l}25\end{tabular}}}}%
    \put(0.60889582,0.29719277){\color[rgb]{0.14901961,0.14901961,0.14901961}\makebox(0,0)[lt]{\lineheight{1.25}\smash{\begin{tabular}[t]{l}30\end{tabular}}}}%
    \put(0.6941912,0.29791966){\color[rgb]{0.14901961,0.14901961,0.14901961}\makebox(0,0)[lt]{\lineheight{1.25}\smash{\begin{tabular}[t]{l}35\end{tabular}}}}%
    \put(0.77976486,0.29719277){\color[rgb]{0.14901961,0.14901961,0.14901961}\makebox(0,0)[lt]{\lineheight{1.25}\smash{\begin{tabular}[t]{l}40\end{tabular}}}}%
    \put(0.03814564,0.33549604){\color[rgb]{0.14901961,0.14901961,0.14901961}\makebox(0,0)[lt]{\lineheight{1.25}\smash{\begin{tabular}[t]{l}-40\end{tabular}}}}%
    \put(0.03814564,0.37893954){\color[rgb]{0.14901961,0.14901961,0.14901961}\makebox(0,0)[lt]{\lineheight{1.25}\smash{\begin{tabular}[t]{l}-30\end{tabular}}}}%
    \put(0.04001561,0.41723656){\color[rgb]{0.14901961,0.14901961,0.14901961}\makebox(0,0)[lt]{\lineheight{1.25}\smash{\begin{tabular}[t]{l}-20\end{tabular}}}}%
    \put(0.03758519,0.4599915){\color[rgb]{0.14901961,0.14901961,0.14901961}\makebox(0,0)[lt]{\lineheight{1.25}\smash{\begin{tabular}[t]{l}-10\end{tabular}}}}%
    \put(0.05789248,0.49621402){\color[rgb]{0.14901961,0.14901961,0.14901961}\makebox(0,0)[lt]{\lineheight{1.25}\smash{\begin{tabular}[t]{l}0\end{tabular}}}}%
    \put(0,0){\includegraphics[width=\unitlength,page=6]{xyzEUC.pdf}}%
    \put(0.86851861,0.5467187){\color[rgb]{0.14901961,0.14901961,0.14901961}\makebox(0,0)[lt]{\lineheight{1.25}\smash{\begin{tabular}[t]{l}45\end{tabular}}}}%
    \put(0.95273461,0.54744554){\color[rgb]{0.14901961,0.14901961,0.14901961}\makebox(0,0)[lt]{\lineheight{1.25}\smash{\begin{tabular}[t]{l}50\end{tabular}}}}%
    \put(0.95368592,0.5166713){\color[rgb]{0.14901961,0.14901961,0.14901961}\makebox(0,0)[lt]{\lineheight{1.25}\smash{\begin{tabular}[t]{l}x10\end{tabular}}}}%
    \put(0,0){\includegraphics[width=\unitlength,page=7]{xyzEUC.pdf}}%
    \put(0.11890574,0.54276121){\color[rgb]{0.14901961,0.14901961,0.14901961}\makebox(0,0)[lt]{\lineheight{1.25}\smash{\begin{tabular}[t]{l}0\end{tabular}}}}%
    \put(0.18045798,0.54306012){\color[rgb]{0.14901961,0.14901961,0.14901961}\makebox(0,0)[lt]{\lineheight{1.25}\smash{\begin{tabular}[t]{l}5\end{tabular}}}}%
    \put(0.26569223,0.54306012){\color[rgb]{0.14901961,0.14901961,0.14901961}\makebox(0,0)[lt]{\lineheight{1.25}\smash{\begin{tabular}[t]{l}10\end{tabular}}}}%
    \put(0.35087434,0.54306012){\color[rgb]{0.14901961,0.14901961,0.14901961}\makebox(0,0)[lt]{\lineheight{1.25}\smash{\begin{tabular}[t]{l}15\end{tabular}}}}%
    \put(0.43683964,0.54306012){\color[rgb]{0.14901961,0.14901961,0.14901961}\makebox(0,0)[lt]{\lineheight{1.25}\smash{\begin{tabular}[t]{l}20\end{tabular}}}}%
    \put(0.52184751,0.54306012){\color[rgb]{0.14901961,0.14901961,0.14901961}\makebox(0,0)[lt]{\lineheight{1.25}\smash{\begin{tabular}[t]{l}25\end{tabular}}}}%
    \put(0.60729062,0.54253876){\color[rgb]{0.14901961,0.14901961,0.14901961}\makebox(0,0)[lt]{\lineheight{1.25}\smash{\begin{tabular}[t]{l}30\end{tabular}}}}%
    \put(0.69258594,0.54326566){\color[rgb]{0.14901961,0.14901961,0.14901961}\makebox(0,0)[lt]{\lineheight{1.25}\smash{\begin{tabular}[t]{l}35\end{tabular}}}}%
    \put(0.77815966,0.54253876){\color[rgb]{0.14901961,0.14901961,0.14901961}\makebox(0,0)[lt]{\lineheight{1.25}\smash{\begin{tabular}[t]{l}40\end{tabular}}}}%
    \put(0.04851191,0.58084202){\color[rgb]{0.14901961,0.14901961,0.14901961}\makebox(0,0)[lt]{\lineheight{1.25}\smash{\begin{tabular}[t]{l}0\end{tabular}}}}%
    \put(0.04452141,0.62428551){\color[rgb]{0.14901961,0.14901961,0.14901961}\makebox(0,0)[lt]{\lineheight{1.25}\smash{\begin{tabular}[t]{l}10\end{tabular}}}}%
    \put(0.04639138,0.66258255){\color[rgb]{0.14901961,0.14901961,0.14901961}\makebox(0,0)[lt]{\lineheight{1.25}\smash{\begin{tabular}[t]{l}20\end{tabular}}}}%
    \put(0.04795146,0.70533749){\color[rgb]{0.14901961,0.14901961,0.14901961}\makebox(0,0)[lt]{\lineheight{1.25}\smash{\begin{tabular}[t]{l}30\end{tabular}}}}%
    \put(0.04830623,0.74156001){\color[rgb]{0.14901961,0.14901961,0.14901961}\makebox(0,0)[lt]{\lineheight{1.25}\smash{\begin{tabular}[t]{l}40\end{tabular}}}}%
    \put(0,0){\includegraphics[width=\unitlength,page=8]{xyzEUC.pdf}}%
  \end{picture}%
\endgroup%

%% file: Figure/TrackingError.pdf_tex
\begingroup%
  \makeatletter%
  \providecommand\color[2][]{%
    \errmessage{(Inkscape) Color is used for the text in Inkscape, but the package 'color.sty' is not loaded}%
    \renewcommand\color[2][]{}%
  }%
  \providecommand\transparent[1]{%
    \errmessage{(Inkscape) Transparency is used (non-zero) for the text in Inkscape, but the package 'transparent.sty' is not loaded}%
    \renewcommand\transparent[1]{}%
  }%
  \providecommand\rotatebox[2]{#2}%
  \newcommand*\fsize{\dimexpr\f@size pt\relax}%
  \newcommand*\lineheight[1]{\fontsize{\fsize}{#1\fsize}\selectfont}%
  \ifx\svgwidth\undefined%
    \setlength{\unitlength}{369.27959599bp}%
    \ifx\svgscale\undefined%
      \relax%
    \else%
      \setlength{\unitlength}{\unitlength * \real{\svgscale}}%
    \fi%
  \else%
    \setlength{\unitlength}{\svgwidth}%
  \fi%
  \global\let\svgwidth\undefined%
  \global\let\svgscale\undefined%
  \makeatother%
  \begin{picture}(1,0.76427178)%
    \lineheight{1}%
    \setlength\tabcolsep{0pt}%
    \put(0.03482442,0.10806494){\color[rgb]{0.14901961,0.14901961,0.14901961}\makebox(0,0)[lt]{\lineheight{1.25}\smash{\begin{tabular}[t]{l}-10\end{tabular}}}}%
    \put(0.0210091,0.63912236){\color[rgb]{0,0,0}\rotatebox{88.30904758}{\makebox(0,0)[lt]{\lineheight{1.25}\smash{\begin{tabular}[t]{l}$e_x \rm{[m]}$\end{tabular}}}}}%
    \put(0.02100908,0.39674448){\color[rgb]{0,0,0}\rotatebox{88.30898061}{\makebox(0,0)[lt]{\lineheight{1.25}\smash{\begin{tabular}[t]{l}$e_y \rm{[m]}$\end{tabular}}}}}%
    \put(0.02038728,0.12864656){\color[rgb]{0,0,0}\rotatebox{88.87783216}{\makebox(0,0)[lt]{\lineheight{1.25}\smash{\begin{tabular}[t]{l}$e_z \rm{[m]}$\end{tabular}}}}}%
    \put(0.5047482,0.00710368){\color[rgb]{0,0,0}\makebox(0,0)[lt]{\lineheight{1.25}\smash{\begin{tabular}[t]{l}Time [s]\end{tabular}}}}%
    \put(0.05357299,0.61280528){\color[rgb]{0.14901961,0.14901961,0.14901961}\makebox(0,0)[lt]{\lineheight{1.25}\smash{\begin{tabular}[t]{l}-50\end{tabular}}}}%
    \put(0,0){\includegraphics[width=\unitlength,page=1]{TrackingError.pdf}}%
    \put(0.12722165,0.57649679){\color[rgb]{0.14901961,0.14901961,0.14901961}\makebox(0,0)[lt]{\lineheight{1.25}\smash{\begin{tabular}[t]{l}0\end{tabular}}}}%
    \put(0.18939448,0.57675223){\color[rgb]{0.14901961,0.14901961,0.14901961}\makebox(0,0)[lt]{\lineheight{1.25}\smash{\begin{tabular}[t]{l}5\end{tabular}}}}%
    \put(0.27548807,0.57675223){\color[rgb]{0.14901961,0.14901961,0.14901961}\makebox(0,0)[lt]{\lineheight{1.25}\smash{\begin{tabular}[t]{l}10\end{tabular}}}}%
    \put(0.36152894,0.57675223){\color[rgb]{0.14901961,0.14901961,0.14901961}\makebox(0,0)[lt]{\lineheight{1.25}\smash{\begin{tabular}[t]{l}15\end{tabular}}}}%
    \put(0.44836096,0.57675223){\color[rgb]{0.14901961,0.14901961,0.14901961}\makebox(0,0)[lt]{\lineheight{1.25}\smash{\begin{tabular}[t]{l}20\end{tabular}}}}%
    \put(0.53422601,0.57675223){\color[rgb]{0.14901961,0.14901961,0.14901961}\makebox(0,0)[lt]{\lineheight{1.25}\smash{\begin{tabular}[t]{l}25\end{tabular}}}}%
    \put(0.62053066,0.5763066){\color[rgb]{0.14901961,0.14901961,0.14901961}\makebox(0,0)[lt]{\lineheight{1.25}\smash{\begin{tabular}[t]{l}30\end{tabular}}}}%
    \put(0.70668589,0.57692795){\color[rgb]{0.14901961,0.14901961,0.14901961}\makebox(0,0)[lt]{\lineheight{1.25}\smash{\begin{tabular}[t]{l}35\end{tabular}}}}%
    \put(0.79312239,0.5763066){\color[rgb]{0.14901961,0.14901961,0.14901961}\makebox(0,0)[lt]{\lineheight{1.25}\smash{\begin{tabular}[t]{l}40\end{tabular}}}}%
    \put(0.87963803,0.57568529){\color[rgb]{0.14901961,0.14901961,0.14901961}\makebox(0,0)[lt]{\lineheight{1.25}\smash{\begin{tabular}[t]{l}45\end{tabular}}}}%
    \put(0.96470302,0.57630656){\color[rgb]{0.14901961,0.14901961,0.14901961}\makebox(0,0)[lt]{\lineheight{1.25}\smash{\begin{tabular}[t]{l}50\end{tabular}}}}%
    \put(0.96566394,0.53781535){\color[rgb]{0.14901961,0.14901961,0.14901961}\makebox(0,0)[lt]{\lineheight{1.25}\smash{\begin{tabular}[t]{l}x10\end{tabular}}}}%
    \put(0.11931858,0.32059309){\color[rgb]{0.14901961,0.14901961,0.14901961}\makebox(0,0)[lt]{\lineheight{1.25}\smash{\begin{tabular}[t]{l}0\end{tabular}}}}%
    \put(0.18149133,0.3208486){\color[rgb]{0.14901961,0.14901961,0.14901961}\makebox(0,0)[lt]{\lineheight{1.25}\smash{\begin{tabular}[t]{l}5\end{tabular}}}}%
    \put(0.26758486,0.3208486){\color[rgb]{0.14901961,0.14901961,0.14901961}\makebox(0,0)[lt]{\lineheight{1.25}\smash{\begin{tabular}[t]{l}10\end{tabular}}}}%
    \put(0.35362566,0.3208486){\color[rgb]{0.14901961,0.14901961,0.14901961}\makebox(0,0)[lt]{\lineheight{1.25}\smash{\begin{tabular}[t]{l}15\end{tabular}}}}%
    \put(0.44045762,0.3208486){\color[rgb]{0.14901961,0.14901961,0.14901961}\makebox(0,0)[lt]{\lineheight{1.25}\smash{\begin{tabular}[t]{l}20\end{tabular}}}}%
    \put(0.52632254,0.3208486){\color[rgb]{0.14901961,0.14901961,0.14901961}\makebox(0,0)[lt]{\lineheight{1.25}\smash{\begin{tabular}[t]{l}25\end{tabular}}}}%
    \put(0.61262722,0.32040297){\color[rgb]{0.14901961,0.14901961,0.14901961}\makebox(0,0)[lt]{\lineheight{1.25}\smash{\begin{tabular}[t]{l}30\end{tabular}}}}%
    \put(0.69878239,0.32102432){\color[rgb]{0.14901961,0.14901961,0.14901961}\makebox(0,0)[lt]{\lineheight{1.25}\smash{\begin{tabular}[t]{l}35\end{tabular}}}}%
    \put(0.78521902,0.32040297){\color[rgb]{0.14901961,0.14901961,0.14901961}\makebox(0,0)[lt]{\lineheight{1.25}\smash{\begin{tabular}[t]{l}40\end{tabular}}}}%
    \put(0.87173459,0.31978159){\color[rgb]{0.14901961,0.14901961,0.14901961}\makebox(0,0)[lt]{\lineheight{1.25}\smash{\begin{tabular}[t]{l}45\end{tabular}}}}%
    \put(0.95679965,0.32040291){\color[rgb]{0.14901961,0.14901961,0.14901961}\makebox(0,0)[lt]{\lineheight{1.25}\smash{\begin{tabular}[t]{l}50\end{tabular}}}}%
    \put(0.95776051,0.28191164){\color[rgb]{0.14901961,0.14901961,0.14901961}\makebox(0,0)[lt]{\lineheight{1.25}\smash{\begin{tabular}[t]{l}x10\end{tabular}}}}%
    \put(0.12919745,0.06265852){\color[rgb]{0.14901961,0.14901961,0.14901961}\makebox(0,0)[lt]{\lineheight{1.25}\smash{\begin{tabular}[t]{l}0\end{tabular}}}}%
    \put(0.19137026,0.06291399){\color[rgb]{0.14901961,0.14901961,0.14901961}\makebox(0,0)[lt]{\lineheight{1.25}\smash{\begin{tabular}[t]{l}5\end{tabular}}}}%
    \put(0.27746388,0.06291399){\color[rgb]{0.14901961,0.14901961,0.14901961}\makebox(0,0)[lt]{\lineheight{1.25}\smash{\begin{tabular}[t]{l}10\end{tabular}}}}%
    \put(0.36350475,0.06291399){\color[rgb]{0.14901961,0.14901961,0.14901961}\makebox(0,0)[lt]{\lineheight{1.25}\smash{\begin{tabular}[t]{l}15\end{tabular}}}}%
    \put(0.45033677,0.06291399){\color[rgb]{0.14901961,0.14901961,0.14901961}\makebox(0,0)[lt]{\lineheight{1.25}\smash{\begin{tabular}[t]{l}20\end{tabular}}}}%
    \put(0.53620173,0.06291399){\color[rgb]{0.14901961,0.14901961,0.14901961}\makebox(0,0)[lt]{\lineheight{1.25}\smash{\begin{tabular}[t]{l}25\end{tabular}}}}%
    \put(0.62250637,0.06246833){\color[rgb]{0.14901961,0.14901961,0.14901961}\makebox(0,0)[lt]{\lineheight{1.25}\smash{\begin{tabular}[t]{l}30\end{tabular}}}}%
    \put(0.7086616,0.06308968){\color[rgb]{0.14901961,0.14901961,0.14901961}\makebox(0,0)[lt]{\lineheight{1.25}\smash{\begin{tabular}[t]{l}35\end{tabular}}}}%
    \put(0.79509811,0.06246833){\color[rgb]{0.14901961,0.14901961,0.14901961}\makebox(0,0)[lt]{\lineheight{1.25}\smash{\begin{tabular}[t]{l}40\end{tabular}}}}%
    \put(0.88161374,0.06184699){\color[rgb]{0.14901961,0.14901961,0.14901961}\makebox(0,0)[lt]{\lineheight{1.25}\smash{\begin{tabular}[t]{l}45\end{tabular}}}}%
    \put(0.96667874,0.06246827){\color[rgb]{0.14901961,0.14901961,0.14901961}\makebox(0,0)[lt]{\lineheight{1.25}\smash{\begin{tabular}[t]{l}50\end{tabular}}}}%
    \put(0.96763966,0.0239771){\color[rgb]{0.14901961,0.14901961,0.14901961}\makebox(0,0)[lt]{\lineheight{1.25}\smash{\begin{tabular}[t]{l}x10\end{tabular}}}}%
    \put(0.04729963,0.45194817){\color[rgb]{0.14901961,0.14901961,0.14901961}\makebox(0,0)[lt]{\lineheight{1.25}\smash{\begin{tabular}[t]{l}-10\end{tabular}}}}%
    \put(0.06620175,0.49548966){\color[rgb]{0.14901961,0.14901961,0.14901961}\makebox(0,0)[lt]{\lineheight{1.25}\smash{\begin{tabular}[t]{l}0\end{tabular}}}}%
    \put(0.04904325,0.41271902){\color[rgb]{0.14901961,0.14901961,0.14901961}\makebox(0,0)[lt]{\lineheight{1.25}\smash{\begin{tabular}[t]{l}-20\end{tabular}}}}%
    \put(0.04904325,0.36917753){\color[rgb]{0.14901961,0.14901961,0.14901961}\makebox(0,0)[lt]{\lineheight{1.25}\smash{\begin{tabular}[t]{l}-30\end{tabular}}}}%
    \put(0.04409804,0.23369208){\color[rgb]{0.14901961,0.14901961,0.14901961}\makebox(0,0)[lt]{\lineheight{1.25}\smash{\begin{tabular}[t]{l}10\end{tabular}}}}%
    \put(0.05609341,0.16870079){\color[rgb]{0.14901961,0.14901961,0.14901961}\makebox(0,0)[lt]{\lineheight{1.25}\smash{\begin{tabular}[t]{l}0\end{tabular}}}}%
    \put(0.05990431,0.75159237){\color[rgb]{0.14901961,0.14901961,0.14901961}\makebox(0,0)[lt]{\lineheight{1.25}\smash{\begin{tabular}[t]{l}50\end{tabular}}}}%
    \put(0.07189968,0.68456995){\color[rgb]{0.14901961,0.14901961,0.14901961}\makebox(0,0)[lt]{\lineheight{1.25}\smash{\begin{tabular}[t]{l}0\end{tabular}}}}%
    \put(0,0){\includegraphics[width=\unitlength,page=2]{TrackingError.pdf}}%
  \end{picture}%
\endgroup%

%% file: Figure/U12.pdf_tex
\begingroup%
  \makeatletter%
  \providecommand\color[2][]{%
    \errmessage{(Inkscape) Color is used for the text in Inkscape, but the package 'color.sty' is not loaded}%
    \renewcommand\color[2][]{}%
  }%
  \providecommand\transparent[1]{%
    \errmessage{(Inkscape) Transparency is used (non-zero) for the text in Inkscape, but the package 'transparent.sty' is not loaded}%
    \renewcommand\transparent[1]{}%
  }%
  \providecommand\rotatebox[2]{#2}%
  \newcommand*\fsize{\dimexpr\f@size pt\relax}%
  \newcommand*\lineheight[1]{\fontsize{\fsize}{#1\fsize}\selectfont}%
  \ifx\svgwidth\undefined%
    \setlength{\unitlength}{335.1977005bp}%
    \ifx\svgscale\undefined%
      \relax%
    \else%
      \setlength{\unitlength}{\unitlength * \real{\svgscale}}%
    \fi%
  \else%
    \setlength{\unitlength}{\svgwidth}%
  \fi%
  \global\let\svgwidth\undefined%
  \global\let\svgscale\undefined%
  \makeatother%
  \begin{picture}(1,1.00333568)%
    \lineheight{1}%
    \setlength\tabcolsep{0pt}%
    \put(0,0){\includegraphics[width=\unitlength,page=1]{U12.pdf}}%
    \put(0.10122014,0.81095296){\color[rgb]{0.14901961,0.14901961,0.14901961}\makebox(0,0)[lt]{\lineheight{1.25}\smash{\begin{tabular}[t]{l}0\end{tabular}}}}%
    \put(0.1643872,0.81123862){\color[rgb]{0.14901961,0.14901961,0.14901961}\makebox(0,0)[lt]{\lineheight{1.25}\smash{\begin{tabular}[t]{l}5\end{tabular}}}}%
    \put(0.25185762,0.81123862){\color[rgb]{0.14901961,0.14901961,0.14901961}\makebox(0,0)[lt]{\lineheight{1.25}\smash{\begin{tabular}[t]{l}10\end{tabular}}}}%
    \put(0.33927446,0.81123862){\color[rgb]{0.14901961,0.14901961,0.14901961}\makebox(0,0)[lt]{\lineheight{1.25}\smash{\begin{tabular}[t]{l}15\end{tabular}}}}%
    \put(0.42749509,0.81123862){\color[rgb]{0.14901961,0.14901961,0.14901961}\makebox(0,0)[lt]{\lineheight{1.25}\smash{\begin{tabular}[t]{l}20\end{tabular}}}}%
    \put(0.51473328,0.81123862){\color[rgb]{0.14901961,0.14901961,0.14901961}\makebox(0,0)[lt]{\lineheight{1.25}\smash{\begin{tabular}[t]{l}25\end{tabular}}}}%
    \put(0.60241807,0.81074036){\color[rgb]{0.14901961,0.14901961,0.14901961}\makebox(0,0)[lt]{\lineheight{1.25}\smash{\begin{tabular}[t]{l}30\end{tabular}}}}%
    \put(0.68995099,0.81143506){\color[rgb]{0.14901961,0.14901961,0.14901961}\makebox(0,0)[lt]{\lineheight{1.25}\smash{\begin{tabular}[t]{l}35\end{tabular}}}}%
    \put(0.77776975,0.81074036){\color[rgb]{0.14901961,0.14901961,0.14901961}\makebox(0,0)[lt]{\lineheight{1.25}\smash{\begin{tabular}[t]{l}40\end{tabular}}}}%
    \put(0.86566887,0.8100457){\color[rgb]{0.14901961,0.14901961,0.14901961}\makebox(0,0)[lt]{\lineheight{1.25}\smash{\begin{tabular}[t]{l}45\end{tabular}}}}%
    \put(0.95209422,0.81074032){\color[rgb]{0.14901961,0.14901961,0.14901961}\makebox(0,0)[lt]{\lineheight{1.25}\smash{\begin{tabular}[t]{l}50\end{tabular}}}}%
    \put(0.95307052,0.75895552){\color[rgb]{0.14901961,0.14901961,0.14901961}\makebox(0,0)[lt]{\lineheight{1.25}\smash{\begin{tabular}[t]{l}x10\end{tabular}}}}%
    \put(0.10122014,0.31665308){\color[rgb]{0.14901961,0.14901961,0.14901961}\makebox(0,0)[lt]{\lineheight{1.25}\smash{\begin{tabular}[t]{l}0\end{tabular}}}}%
    \put(0.1643872,0.31693869){\color[rgb]{0.14901961,0.14901961,0.14901961}\makebox(0,0)[lt]{\lineheight{1.25}\smash{\begin{tabular}[t]{l}5\end{tabular}}}}%
    \put(0.25185757,0.31693869){\color[rgb]{0.14901961,0.14901961,0.14901961}\makebox(0,0)[lt]{\lineheight{1.25}\smash{\begin{tabular}[t]{l}10\end{tabular}}}}%
    \put(0.33927446,0.31693869){\color[rgb]{0.14901961,0.14901961,0.14901961}\makebox(0,0)[lt]{\lineheight{1.25}\smash{\begin{tabular}[t]{l}15\end{tabular}}}}%
    \put(0.42749509,0.31693869){\color[rgb]{0.14901961,0.14901961,0.14901961}\makebox(0,0)[lt]{\lineheight{1.25}\smash{\begin{tabular}[t]{l}20\end{tabular}}}}%
    \put(0.51473325,0.31693869){\color[rgb]{0.14901961,0.14901961,0.14901961}\makebox(0,0)[lt]{\lineheight{1.25}\smash{\begin{tabular}[t]{l}25\end{tabular}}}}%
    \put(0.60241807,0.31644048){\color[rgb]{0.14901961,0.14901961,0.14901961}\makebox(0,0)[lt]{\lineheight{1.25}\smash{\begin{tabular}[t]{l}30\end{tabular}}}}%
    \put(0.68995099,0.31713517){\color[rgb]{0.14901961,0.14901961,0.14901961}\makebox(0,0)[lt]{\lineheight{1.25}\smash{\begin{tabular}[t]{l}35\end{tabular}}}}%
    \put(0.77776975,0.31644048){\color[rgb]{0.14901961,0.14901961,0.14901961}\makebox(0,0)[lt]{\lineheight{1.25}\smash{\begin{tabular}[t]{l}40\end{tabular}}}}%
    \put(0.8656688,0.31574593){\color[rgb]{0.14901961,0.14901961,0.14901961}\makebox(0,0)[lt]{\lineheight{1.25}\smash{\begin{tabular}[t]{l}45\end{tabular}}}}%
    \put(0.95209422,0.31644041){\color[rgb]{0.14901961,0.14901961,0.14901961}\makebox(0,0)[lt]{\lineheight{1.25}\smash{\begin{tabular}[t]{l}50\end{tabular}}}}%
    \put(0.95708526,0.26465567){\color[rgb]{0.14901961,0.14901961,0.14901961}\makebox(0,0)[lt]{\lineheight{1.25}\smash{\begin{tabular}[t]{l}x10\end{tabular}}}}%
    \put(0.09720539,0.55855949){\color[rgb]{0.14901961,0.14901961,0.14901961}\makebox(0,0)[lt]{\lineheight{1.25}\smash{\begin{tabular}[t]{l}0\end{tabular}}}}%
    \put(0.16037244,0.55884513){\color[rgb]{0.14901961,0.14901961,0.14901961}\makebox(0,0)[lt]{\lineheight{1.25}\smash{\begin{tabular}[t]{l}5\end{tabular}}}}%
    \put(0.24784281,0.55884513){\color[rgb]{0.14901961,0.14901961,0.14901961}\makebox(0,0)[lt]{\lineheight{1.25}\smash{\begin{tabular}[t]{l}10\end{tabular}}}}%
    \put(0.33525968,0.55884513){\color[rgb]{0.14901961,0.14901961,0.14901961}\makebox(0,0)[lt]{\lineheight{1.25}\smash{\begin{tabular}[t]{l}15\end{tabular}}}}%
    \put(0.42348035,0.55884513){\color[rgb]{0.14901961,0.14901961,0.14901961}\makebox(0,0)[lt]{\lineheight{1.25}\smash{\begin{tabular}[t]{l}20\end{tabular}}}}%
    \put(0.51071847,0.55884513){\color[rgb]{0.14901961,0.14901961,0.14901961}\makebox(0,0)[lt]{\lineheight{1.25}\smash{\begin{tabular}[t]{l}25\end{tabular}}}}%
    \put(0.59840333,0.55834685){\color[rgb]{0.14901961,0.14901961,0.14901961}\makebox(0,0)[lt]{\lineheight{1.25}\smash{\begin{tabular}[t]{l}30\end{tabular}}}}%
    \put(0.68593624,0.55904154){\color[rgb]{0.14901961,0.14901961,0.14901961}\makebox(0,0)[lt]{\lineheight{1.25}\smash{\begin{tabular}[t]{l}35\end{tabular}}}}%
    \put(0.773755,0.55834685){\color[rgb]{0.14901961,0.14901961,0.14901961}\makebox(0,0)[lt]{\lineheight{1.25}\smash{\begin{tabular}[t]{l}40\end{tabular}}}}%
    \put(0.86165412,0.55765223){\color[rgb]{0.14901961,0.14901961,0.14901961}\makebox(0,0)[lt]{\lineheight{1.25}\smash{\begin{tabular}[t]{l}45\end{tabular}}}}%
    \put(0.94807948,0.55834682){\color[rgb]{0.14901961,0.14901961,0.14901961}\makebox(0,0)[lt]{\lineheight{1.25}\smash{\begin{tabular}[t]{l}50\end{tabular}}}}%
    \put(0.94905571,0.506562){\color[rgb]{0.14901961,0.14901961,0.14901961}\makebox(0,0)[lt]{\lineheight{1.25}\smash{\begin{tabular}[t]{l}x10\end{tabular}}}}%
    \put(0.11527181,0.05587003){\color[rgb]{0.14901961,0.14901961,0.14901961}\makebox(0,0)[lt]{\lineheight{1.25}\smash{\begin{tabular}[t]{l}0\end{tabular}}}}%
    \put(0.17843887,0.05615571){\color[rgb]{0.14901961,0.14901961,0.14901961}\makebox(0,0)[lt]{\lineheight{1.25}\smash{\begin{tabular}[t]{l}5\end{tabular}}}}%
    \put(0.2659093,0.05615571){\color[rgb]{0.14901961,0.14901961,0.14901961}\makebox(0,0)[lt]{\lineheight{1.25}\smash{\begin{tabular}[t]{l}10\end{tabular}}}}%
    \put(0.35332613,0.05615571){\color[rgb]{0.14901961,0.14901961,0.14901961}\makebox(0,0)[lt]{\lineheight{1.25}\smash{\begin{tabular}[t]{l}15\end{tabular}}}}%
    \put(0.44154676,0.05615571){\color[rgb]{0.14901961,0.14901961,0.14901961}\makebox(0,0)[lt]{\lineheight{1.25}\smash{\begin{tabular}[t]{l}20\end{tabular}}}}%
    \put(0.52878495,0.05615571){\color[rgb]{0.14901961,0.14901961,0.14901961}\makebox(0,0)[lt]{\lineheight{1.25}\smash{\begin{tabular}[t]{l}45\end{tabular}}}}%
    \put(0.61646974,0.05565743){\color[rgb]{0.14901961,0.14901961,0.14901961}\makebox(0,0)[lt]{\lineheight{1.25}\smash{\begin{tabular}[t]{l}30\end{tabular}}}}%
    \put(0.70400266,0.05635212){\color[rgb]{0.14901961,0.14901961,0.14901961}\makebox(0,0)[lt]{\lineheight{1.25}\smash{\begin{tabular}[t]{l}35\end{tabular}}}}%
    \put(0.79182142,0.05565743){\color[rgb]{0.14901961,0.14901961,0.14901961}\makebox(0,0)[lt]{\lineheight{1.25}\smash{\begin{tabular}[t]{l}40\end{tabular}}}}%
    \put(0.87972054,0.05496281){\color[rgb]{0.14901961,0.14901961,0.14901961}\makebox(0,0)[lt]{\lineheight{1.25}\smash{\begin{tabular}[t]{l}45\end{tabular}}}}%
    \put(0.96614596,0.05565736){\color[rgb]{0.14901961,0.14901961,0.14901961}\makebox(0,0)[lt]{\lineheight{1.25}\smash{\begin{tabular}[t]{l}50\end{tabular}}}}%
    \put(0.96712219,0.01729752){\color[rgb]{0.14901961,0.14901961,0.14901961}\makebox(0,0)[lt]{\lineheight{1.25}\smash{\begin{tabular}[t]{l}x10\end{tabular}}}}%
    \put(0.04252724,0.44372691){\color[rgb]{0.14901961,0.14901961,0.14901961}\makebox(0,0)[lt]{\lineheight{1.25}\smash{\begin{tabular}[t]{l}10\end{tabular}}}}%
    \put(0.04165783,0.48786588){\color[rgb]{0.14901961,0.14901961,0.14901961}\makebox(0,0)[lt]{\lineheight{1.25}\smash{\begin{tabular}[t]{l}15\end{tabular}}}}%
    \put(0.05232827,0.39986794){\color[rgb]{0.14901961,0.14901961,0.14901961}\makebox(0,0)[lt]{\lineheight{1.25}\smash{\begin{tabular}[t]{l}5\end{tabular}}}}%
    \put(0.05232827,0.35572904){\color[rgb]{0.14901961,0.14901961,0.14901961}\makebox(0,0)[lt]{\lineheight{1.25}\smash{\begin{tabular}[t]{l}0\end{tabular}}}}%
    \put(0.04093029,0.94525417){\color[rgb]{0.14901961,0.14901961,0.14901961}\makebox(0,0)[lt]{\lineheight{1.25}\smash{\begin{tabular}[t]{l}10\end{tabular}}}}%
    \put(0.04006088,0.98939313){\color[rgb]{0.14901961,0.14901961,0.14901961}\makebox(0,0)[lt]{\lineheight{1.25}\smash{\begin{tabular}[t]{l}15\end{tabular}}}}%
    \put(0.04671656,0.90139519){\color[rgb]{0.14901961,0.14901961,0.14901961}\makebox(0,0)[lt]{\lineheight{1.25}\smash{\begin{tabular}[t]{l}5\end{tabular}}}}%
    \put(0.04671656,0.85725622){\color[rgb]{0.14901961,0.14901961,0.14901961}\makebox(0,0)[lt]{\lineheight{1.25}\smash{\begin{tabular}[t]{l}0\end{tabular}}}}%
    \put(0.04156459,0.69279834){\color[rgb]{0.14901961,0.14901961,0.14901961}\makebox(0,0)[lt]{\lineheight{1.25}\smash{\begin{tabular}[t]{l}10\end{tabular}}}}%
    \put(0.04069518,0.73693729){\color[rgb]{0.14901961,0.14901961,0.14901961}\makebox(0,0)[lt]{\lineheight{1.25}\smash{\begin{tabular}[t]{l}15\end{tabular}}}}%
    \put(0.05136563,0.64893933){\color[rgb]{0.14901961,0.14901961,0.14901961}\makebox(0,0)[lt]{\lineheight{1.25}\smash{\begin{tabular}[t]{l}5\end{tabular}}}}%
    \put(0.05136563,0.60480036){\color[rgb]{0.14901961,0.14901961,0.14901961}\makebox(0,0)[lt]{\lineheight{1.25}\smash{\begin{tabular}[t]{l}0\end{tabular}}}}%
    \put(0.04492266,0.19077163){\color[rgb]{0.14901961,0.14901961,0.14901961}\makebox(0,0)[lt]{\lineheight{1.25}\smash{\begin{tabular}[t]{l}10\end{tabular}}}}%
    \put(0.04405326,0.2349106){\color[rgb]{0.14901961,0.14901961,0.14901961}\makebox(0,0)[lt]{\lineheight{1.25}\smash{\begin{tabular}[t]{l}15\end{tabular}}}}%
    \put(0.0547237,0.14691262){\color[rgb]{0.14901961,0.14901961,0.14901961}\makebox(0,0)[lt]{\lineheight{1.25}\smash{\begin{tabular}[t]{l}5\end{tabular}}}}%
    \put(0.0547237,0.10277366){\color[rgb]{0.14901961,0.14901961,0.14901961}\makebox(0,0)[lt]{\lineheight{1.25}\smash{\begin{tabular}[t]{l}0\end{tabular}}}}%
    \put(0,0){\includegraphics[width=\unitlength,page=2]{U12.pdf}}%
    \put(0.71741055,0.96051665){\color[rgb]{0,0,0}\makebox(0,0)[lt]{\lineheight{1.25}\smash{\begin{tabular}[t]{l}$artificial\;input$\end{tabular}}}}%
    \put(0.71741055,0.923382){\color[rgb]{0,0,0}\makebox(0,0)[lt]{\lineheight{1.25}\smash{\begin{tabular}[t]{l}$input$\end{tabular}}}}%
    \put(0.47998578,0.00794205){\color[rgb]{0,0,0}\makebox(0,0)[lt]{\lineheight{1.25}\smash{\begin{tabular}[t]{l}Time [s]\end{tabular}}}}%
    \put(0.02230155,0.60526703){\color[rgb]{0,0,0}\rotatebox{87.97388509}{\makebox(0,0)[lt]{\lineheight{1.25}\smash{\begin{tabular}[t]{l}${f}_2 \, \si{[\newton]}$\end{tabular}}}}}%
    \put(0.01929969,0.86707704){\color[rgb]{0,0,0}\rotatebox{86.63795269}{\makebox(0,0)[lt]{\lineheight{1.25}\smash{\begin{tabular}[t]{l}${f}_1 \, \si{[\newton]}$\end{tabular}}}}}%
    \put(0.02247004,0.39946498){\color[rgb]{0,0,0}\rotatebox{89.49124105}{\makebox(0,0)[lt]{\lineheight{1.25}\smash{\begin{tabular}[t]{l}${f}_3 \, \si{[\newton]}$\end{tabular}}}}}%
    \put(0.0176802,0.14510676){\color[rgb]{0,0,0}\rotatebox{88.26890379}{\makebox(0,0)[lt]{\lineheight{1.25}\smash{\begin{tabular}[t]{l}${f}_4 \, \si{[\newton]}$\end{tabular}}}}}%
    \put(0,0){\includegraphics[width=\unitlength,page=3]{U12.pdf}}%
  \end{picture}%
\endgroup%